\documentclass[a4paper,11pt]{article}
\pdfoutput=1

\usepackage{jcappub}

\usepackage[T1]{fontenc}
\usepackage{bm}

\newcommand{\mfp}{\lambda_{912}^{\mathrm{mfp}}}
\newcommand{\HI}{H{\sc~i}}

\title{\boldmath Small-scale clumping of dark matter and the mean free path of ionizing photons at $z=6$}

\author[a]{Christopher Cain,\note{Corresponding author.}}
\author[a]{Anson D'Aloisio,}
\author[b,c]{Vid Ir\v{s}i\v{c},}
\author[a]{Nakul Gangolli,}
\author[a]{Sanya Dhami}

\affiliation[a]{Department of Physics and Astronomy, University of California, 900 University Ave, Riverside, CA 92521, USA}
\affiliation[b]{Kavli Institute for Cosmology, University of Cambridge, Madingley Road, Cambridge CB3 0HA, UK}
\affiliation[c]{Cavendish Laboratory, University of Cambridge, 19 J. J. Thomson Ave, Cambridge CB3 0HE, UK}

\emailAdd{ccain002@ucr.edu}

\abstract{Recently, the mean free path of ionizing photons in the $z = 6$ intergalactic medium (IGM) was measured to be very short, presenting a challenge to existing reionization models.  At face value, the measurement can be interpreted as evidence that the IGM clumps on scales $M\lesssim 10^8$ M$_\odot$, a key but largely untested prediction of the cold dark matter (CDM) paradigm.  Motivated by this possibility, we study the role that the underlying dark matter cosmology plays in setting the $z > 5$ mean free path. We use two classes of models to contrast against the standard CDM prediction: (1) thermal relic warm dark matter (WDM), representing models with suppressed small-scale power; (2) an ultralight axion exhibiting a white noise-like power enhancement.   Differences in the mean free path between the WDM and CDM models are subdued by pressure smoothing and the possible contribution of neutral islands to the IGM opacity.  For example, comparing late reionization scenarios with a fixed volume-weighted mean neutral fraction of $20\%$ at $z=6$, the mean free path is $19~(45)~\%$ longer in a WDM model with $m_x = 3~(1)$ keV.    The enhanced power in the axion-like model produces better agreement with the short mean free path measured at $z = 6$. However, drawing robust conclusions about cosmology is hampered by large uncertainties in the reionization process, extragalactic ionizing background, and thermal history of the Universe.  This work highlights some key open questions about the IGM opacity during reionization.}

\begin{document}
\maketitle
\flushbottom

\section{Introduction} \label{sec:intro}

\begin{figure}
    \centering
\resizebox{14cm}{!}{\includegraphics[trim={0cm 0cm 0cm 0cm},clip]{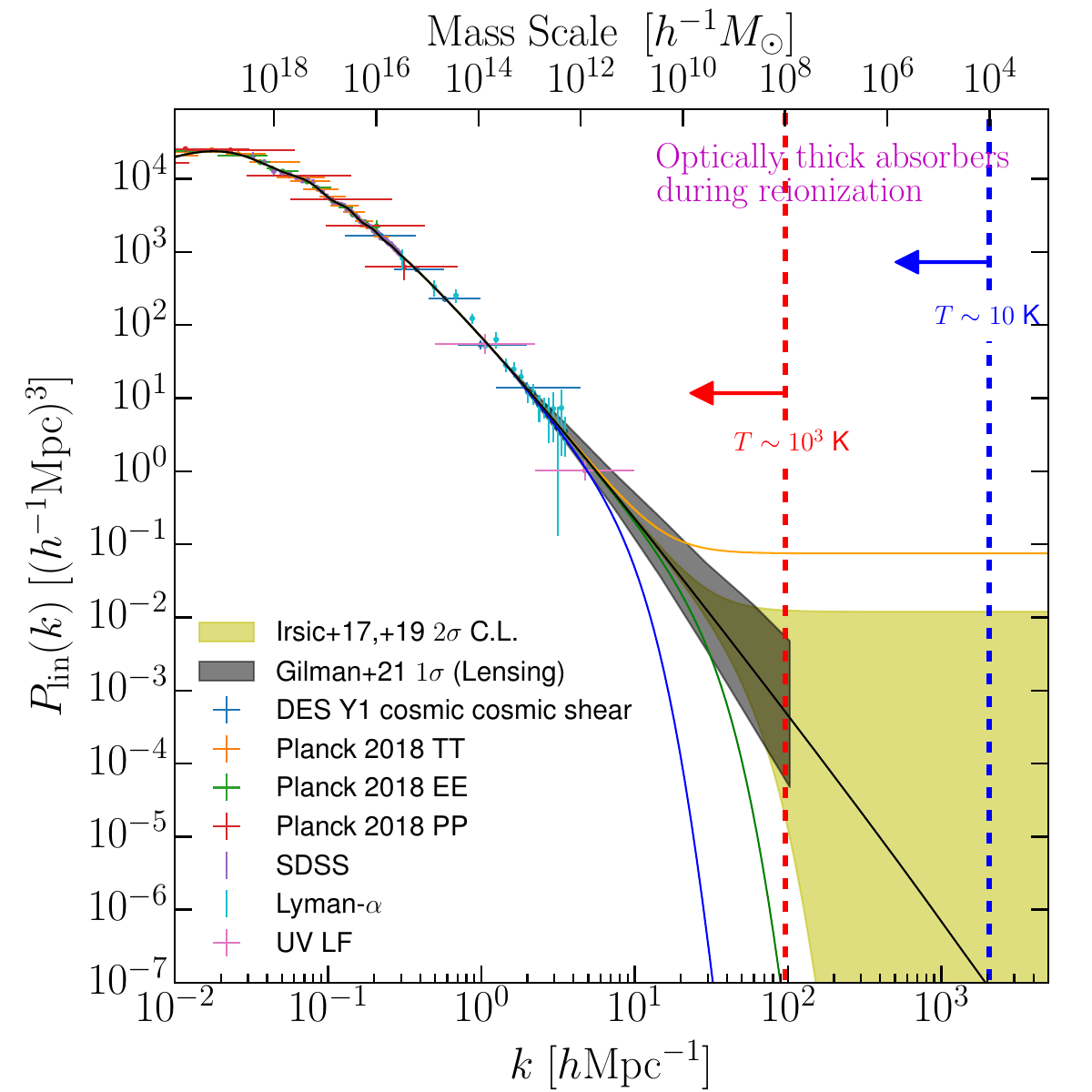}}
    \caption{ Summary of constraints on the linearly extrapolated matter power spectrum.  Power on the largest scales is constrained by Planck CMB measurements \cite{Planck2018} and galaxy clustering in the Sloan Digital Sky Survey (SDSS)~\cite{Reid2010}.  On intermediate scales, constraints come from cosmic shear  measurements in the Dark Energy Survey (DES)~\cite{Troxel2018}, the Lyman-$\alpha$ forest 1D flux power spectrum ~\cite{Viel2004,Chabanier2019}, and the UV luminosity function of high-$z$ galaxies \cite{2021arXiv211013161S}.  The thin curves show the different DM models considered in this work.  The black curve is the concordance CDM power spectrum.  The yellow, green and blue curves show an axion-like cosmology with enhanced small scale power, and thermal relic WDM scenarios with masses $m_X = 3$ and $1$ keV, respectively.  The gray shaded region denotes the $1\sigma$ constraints from~\cite{Gilman2021}, derived from flux ratios and positions of strongly lensed quasars.  The yellow-shaded region denotes the range spanned by the $2\sigma$ lower limits on the thermal relic WDM particle mass from \cite{Irsic2017} and the 2$\sigma$ upper limits on the isocurvature fraction in the ultra-light axion scenario from \cite{Irsic2019}.  Although much of the constraining power from the Ly$\alpha$ forest and lensing comes from mass scales as small as $M=10^{8}$ $h^{-1}$M$_\odot$, on smaller scales (right of the red dashed line) the constraints from Refs~\cite{Irsic2017,Irsic2019} are extrapolations of the assumed DM cosmology and are thus strongly model-dependent.  As such these scales are effectively unconstrained.  The thick vertical dashed lines denote rough lower limits on the range of mass scales expected to contribute to the opacity of the IGM during reionization for two different minimum pre-reionization gas temperatures. The IGM opacity during and shortly after ($\Delta t \sim 300$ Myr) reionization might have been sensitive to power on scales yet unconstrained. The possibility of exploiting this to constrain DM models is the topic of this paper.  }
    \label{fig:powerspec}
\end{figure}

Small-scale power is a defining feature of cold collisionless dark matter (CDM), manifested in halo formation down to perhaps Earth-mass scales \cite{2005PhRvD..71j3520L, 2006PhRvL..97c1301P, 2006PhRvD..74f3509B, 2006ApJ...649....1D}.  Figure \ref{fig:powerspec} summarizes some recent constraints on the linear matter power spectrum, $P_{\rm lin}(k)$, across the range of scales currently accessible to observations.  The top horizontal axis shows the Larangian mass scale corresponding to wavenumber $k$, $M = \frac{4 \pi}{3}\rho_{m}(z = 0) R^3$, where $R = 2\pi/k$ and $\rho_m$ is the cosmological matter density.  The Lyman-$\alpha$ forest flux power spectrum is sensitive to $P_{\rm lin}(k)$ up to wavenumber $k\approx 50~h$Mpc$^{-1}$, with the most recent measurements placing tight limits on the parameter space of CDM alternatives \cite{Viel2013, Palanque-Delabrouille2013, Irsic2017, 2017PhRvL.119c1302I, 2017JCAP...12..013B, Murgia2018, Chabanier2019}\footnote{We emphasize that the gold and gray shaded regions in Fig. \ref{fig:powerspec} correspond to {\it model dependent} constraints on $P_{\rm lin}(k)$ .}.    Probing $P_{\rm lin}(k)$ on smaller scales, flux ratio measurements in strong gravitational lenses are currently sensitive to perturbations by halos with masses $M\gtrsim 2 \times10^{7}$ $h^{-1}$M$_\odot$, corresponding to $k\approx 160~h$Mpc$^{-1}$ \cite{2019MNRAS.487.5721G,Gilman2021,2022MNRAS.512.5862H}.  Future observations by the James Webb Space Telescope (JWST) aim to extend this sensitivity to $M\sim 2\times 10^{6}$ $h^{-1}$M$_\odot$, or $k\approx 340$ $h{\rm Mpc}^{-1}$ (JWST GO-02046; PI Nierenberg). Developing methods to measure power on even smaller scales is of great interest, given its status as an inevitable but largely untested feature of CDM, and for its potential in probing inflationary physics.

Observations of the abundance and properties of Milky Way satellites can in principle constrain $P_{\rm lin}(k)$ down to the scales of the smallest observable galaxies -- a program termed near-field cosmology.  In theory, halos with masses much below $10^8$ M$_\odot$ are thought to be extremely inefficient at forming stars because they lack a robust cooling channel to kickstart star formation\footnote{See however Refs \cite{Wise2014,Norman2018}. }.  They also struggle to accrete/retain gas against feedback and heating by the extragalactic ionizing background \cite{1994ApJ...427...25S, 1996ApJ...465..608T, 2006MNRAS.371..401H, 2008MNRAS.390..920O, 2013MNRAS.432L..51S, 2014MNRAS.444..503N}.  According to simulations, dense remnants of at least a significant fraction of these barren halos are expected to survive to the present day, even after they are incorporated into larger halos  \cite[e.g.][]{2018MNRAS.474.3043V}.  Strong gravitational lensing magnifications are uniquely sensitive to such dark matter-dominated structures.  Together with near-field cosmology, the forthcoming expansion of strong lensing observations will play a chief role in nailing down the small-scale $P_{\rm lin}(k)$ \cite{2015ApJ...811...20C}.  However, inverting these observations still requires an accurate model connecting the abundance and properties of surviving halos to $P_{\rm lin}(k)$ -- an extremely formidable challenge.  It is therefore important to explore complimentary probes moving forward. 

Reionization-era observations could, at least in principle, give insights into $P_{\rm lin}(k)$ from a much earlier time in the hierarchical assembly process.  In fact, a census of reionization-era galaxies has already been applied to constrain $P_{\rm lin}(k)$ up to $k\sim 10~h$Mpc$^{-1}$ (see Fig. \ref{fig:powerspec}), albeit with large uncertainty \cite{2021arXiv211013161S}.  Another window, as of yet unexploited, comes from the fact that reionization was also shaped by the small-scale structure of the IGM.  In the cold pre-reionization IGM, the Jeans filtering scale was much smaller than it is today, somewhere in the range of $M\sim10^4 - 10^8$ $h^{-1}$M$_{\odot}$, corresponding to characteristic temperatures of $T\sim 10 - 1,000$ K. These scales are denoted by the vertical dashed lines in Fig. \ref{fig:powerspec}. The quoted filtering scales bracket roughly the range of existing models for heating of the IGM by the first X-ray sources, preceding reionization \cite{Ripamonti2008, Jeon2013}.

It has long been recognized that the ``minihalos'' with masses above the filtering scale, but below the minimum mass-scale for efficient galaxy formation, may have contributed significantly to the Lyman-continuum (LyC) opacity of the IGM during reionization \cite{Shapiro2004,Iliev2005}. Absorptions by these halos raised the ionizing photon budget required to complete and maintain reionization, and possibly set the shapes and sizes of ionized bubbles \cite{2005ApJ...624..491I,McQuinn2007}.  The minihalos did not retain their gas content indefinitely, however. They were evacuated over a timescale of $\Delta t \sim 10$ to a few hundred Myr as ionization fronts (I-fronts) eventually penetrated inward, driving evaporative winds into the IGM.  Especially in the earlier stages of reionization, a significant number of absorptions likely occurred outside of halos as well, owing to the higher cosmic densities and weaker ionizing background \cite{Nasir2021}.\footnote{In fact, the results of \cite{Nasir2021} suggest that the diffuse inter-halo gas contributed a LyC opacity roughly equal to that of the minihalos before the latter were photoevaporated.}  After I-fronts passed through a region, pressure smoothing caused the inter-halo gas, e.g. within filaments, to expand outward and relax to a more diffuse configuration within a few hundred Myr.  Simulating these dynamic processes requires hydrodynamics coupled with radiative transfer to capture the interplay between self-shielding and the response of the gas to photoheating \cite{Park2016, DAloisio2020}.  But the physics is comparatively simpler than the highly uncertain processes that shaped the properties of star-forming galaxies and their subhalos. Assuming that the gas structure of the IGM was not significantly spoiled by feedback from the star formation activity of halos, the sinks of reionization could potentially become a useful test for the existence of small-scale power.

Measuring the LyC mean free path of the IGM during reionization is the most direct way to probe the sinks.  The highest redshift constraints to date were reported recently by Refs \cite{Becker2021} and \cite{Bosman2021b} using quasar absorption spectra at $z\approx 6$.     According to recent models which place the end of reionization at around $z=5.2$ \cite{Kulkarni2019, Keating2019, Nasir2020}, the measurement of \cite{Becker2021} might be the first direct measurement of the mean free path during reionization, at a time when the global neutral fraction was $\approx 10\%$. Their measurement of $\mfp(z=6) = 3.57^{+3.09}_{-2.14}$ $h^{-1}$cMpc came as somewhat of a surprise, though, because it is significantly shorter than predictions from the contemporaneous simulations of reionization.  Shortly afterward,  Ref \cite{Cain2021} was able to recover values of $\mfp(z=6)$ compatible with the measurement.  Crucially, their simulations included a sub-grid model of the sinks based on the highly resolved radiative hydrodynamics simulations of \cite{DAloisio2020}.  Thus, Ref \cite{Cain2021} was able to incorporate the effects of small-scale power down to $\sim 10h^{-1}$ckpc scales in reionization simulations with box size $L=200 h^{-1}$cMpc. The key point is that the highly resolved simulations upon which their sub-grid model is based contain tiny gaseous structures close to the lower limit of filtering scales quoted above, $\sim 10^4$ M$_{\odot}$ (see \cite{Nasir2021} for a detailed discussion).\footnote{This owes to the fact that the simulations did not include any pre-heating by X-ray sources ahead of reionization.}  The apparent necessity of including such small structures to reproduce the short value of $\mfp(z=6)$, if correct, suggests a potential broader implication for cosmology. Might the small-scale power predicted in the CDM paradigm be necessary to explain the LyC opacity of the reionizing IGM?   This is among the central questions that we attempt to address here.

In this paper, we examine the connection between the small-scale clumping of the underlying dark matter model and the observed mean free path of the IGM at $z\gtrsim 5$. We employ hydrodynamic simulations in warm dark matter (WDM) cosmologies to quantify the mean free path in models with a small-scale cutoff in $P_{\rm lin}(k)$. We also use a semi-analytic approach to explore scenarios with enhanced small-scale power relative to CDM.  This is motivated by models receiving increased interest in recent years, e.g. ultralight axion-like particles \cite{Preskill1983, Efstathiou1986, Hogan1988, Irsic2019, Dai2020, Xiao2021} and primordial black holes \cite{Afshordi2003,Frampton2010,Belotsky2014,Bird2016b,Clesse2017}, and by the fact that the observed  $\mfp(z=6) = 3.57^{+3.09}_{-2.14}$ $h^{-1}$cMpc lies on the shorter side of expectation in the standard CDM picture~\cite{Becker2021,Cain2021,Davies2021}. The impetus for this exploration was the possibility that high-$z$ mean free path measurements could become a kind of ``no-go'' test for all dark matter models lacking in small scale power.   Unfortunately, as we will show, the situation turns out to be more complicated because of substantial uncertainties in the reionization process, the extragalactic ionizing background, and thermal history of the IGM. 

This work is organized as follows.  \S\ref{sec:numerical} describes our modeling methods. \S\ref{sec:results} discusses the mean free path in WDM models. \S\ref{sec:enhancedpk} presents our axion-like scenario with enhanced small-scale power. In \S\ref{sec:conc} we offer concluding remarks.  Throughout this work, we assume the following cosmological parameters: $\Omega_m = 0.305$, $\Omega_{\Lambda} = 1 - \Omega_m$, $\Omega_b = 0.048$, $h = 0.68$, $n_s = 0.9667$ and $\sigma_8 = 0.82$, consistent with the latest constraints \cite{Planck2018}.  

\section{Numerical Methods} \label{sec:numerical}

\subsection{Hydrodynamic simulations of the sinks} \label{subsec:densPDF}

We ran high-resolution hydrodynamic simulations of the IGM in CDM and WDM scenarios. (We will describe our semi-analytic approach for modeling enhanced power scenarios in \S \ref{sec:enhancedpk}.) We used a modified version of the RadHydro code~\cite{Trac2004} in boxes with $L=2h^{-1}$Mpc, initialized at $z = 300$ using transfer functions generated with CAMB \cite{Lewis2000}.  WDM cosmologies were implemented using the standard approach of Refs \cite{Bode2001,Viel2005}.  We consider thermal relic WDM with particle masses of $m_X = 1$ and 3 keV.\footnote{When quoting DM particle masses, we will use the standard convention of setting $c=1$.}  The former was chosen to be an extreme case which is already ruled out observationally, while the latter is representative of models marginally allowed by recent Ly$\alpha$ forest analyses \cite{Irsic2017}.  The blue and green curves in Figure~\ref{fig:powerspec} show the corresponding linear matter power spectra extrapolated to $z=0$. Structure is suppressed below the free-streaming scale, which is (for thermal relic WDM) $k_F\sim 15$ and  $45~h$Mpc$^{-1}$ for $m_X = 1$ and 3 keV, respectively (Eq. 8 of Ref \cite{Viel2005}).

As described in \S \ref{sec:intro}, pressure smoothing is a key ingredient for modeling the sinks \cite{Park2016, DAloisio2020, Nasir2021}.  To incorporate these effects, we ran simulations with and without a uniform ionizing background applied.   For the former, the ionizing background was switched on at $z=12$, with intensity fixed to a hydrogen photoionization rate of $\Gamma_{-12} \equiv \Gamma_{\rm HI}/10^{-12}$ s$^{-1}= 0.3$ at $z > 6$.  For $z<6$, $\Gamma_{\rm HI}$ evolves to approximately match observational measurements from the Ly$\alpha$ forest.  The evolution of $\Gamma_{\rm HI}$ in our simulations is plotted, along with recent forest constraints, in \S \ref{subsec:astro}. We account for self-shielding with a model calibrated to the fully coupled radiative hydrodynamics simulations of \cite{DAloisio2020, Nasir2021}.   Details are described in Appendix \ref{app:selfshielding}.  In summary, we use a modified version of the fitting function of \cite{Rahmati2013} for the photoionization rate as a function of the local hydrogen density.  We modified the functional form and fitting parameters to match the median $\Gamma_{\rm HI}(n_H)$ reported by \cite{Nasir2021} (see their Fig. 1).  The intensity of the ionizing background declines steeply within density peaks, mimicking the effects of self-shielding observed in radiative hydrodynamics simulations.  Note that our basic simulation setup and code is the same as that of \cite{DAloisio2020}, the main differences being the implementation of WDM, the use of a self-shielding prescription in lieu of full RT, and the evolution of $\Gamma_{\rm HI}$ in the relaxed runs.

With this setup, the gas in the runs that apply an ionizing background is almost instantaneously heated to $T\sim 20,000$ K at $z=12$.  The purpose of this early, impulsive heating is to achieve a limiting case in which photoionization heating has had sufficient time ($\Delta t \gtrsim 300$ Myr) to smooth the density structure of the IGM by $z=6$. We will refer to this smoothing process as ``relaxation,'' and label the corresponding runs as ``relaxed.''   For a fixed ionizing background intensity ($\Gamma_{\rm HI}$), most of the relaxation and photo-evaporation occurs within a time $\Delta t \sim 300$ Myr since I-front passage, after which the local density field possess essentially no memory of when it was reionized.  Hence the density structures at $z\lesssim 6$ in our relaxed runs are representative of IGM patches that were ionized/heated during the first half of reionization (up to differences in the local $\Gamma_{\rm HI}$, which we will address below). In the opposite limit, no ionizing background was applied, such that the cold gas clumps down to its pre-reionization Jeans filtering scale, more representative of very recently reionized patches (see e.g. \cite{Emberson2013}).  We refer to these runs as ``un-relaxed.''   We will use a simple model for the evolution connecting these two limiting configurations, described in \S \ref{subsec:reion_model}. 

In Appendix~\ref{app:convergence}, we show that numerical convergence can be achieved with a smaller number of DM particles ($N_{\rm dm}$) and gas cells ($N_{\rm gas}$) in the relaxed runs, compared to the un-relaxed ones.  This finding reflects the different filtering scales, or minimum sizes of gaseous structures to form, in the relaxed and un-relaxed runs. A similar effect applies to the free streaming scale; the lighter the WDM particle, the less stringent the resolution requirements.  We adopted $N \equiv N_{\rm dm} = N_{\rm gas} = 2048^3$  for the CDM and $m_X = 3$ keV un-relaxed runs, and $N=1024^3$ for the corresponding relaxed runs.  We used $N = 1024^3$ for both the un-relaxed and relaxed runs with $m_X = 1$ keV.   In Appendix~\ref{app:convergence}, we justify the use of a lower resolution for the relaxed and $m_X = 1$ keV runs.  We also show that, in the un-relaxed limit, the WDM runs are better converged than the CDM runs, owing to the intrinsic lack of small-scale power in the former. As a result, we likely underestimate the differences in opacities between our WDM and CDM models in the un-relaxed limit.  Note, however, that our simulations do not include the effects of pre-heating by the first X-ray sources.  The pre-heating would act in the opposite direction, smoothing out the smallest gaseous structures present in the CDM cosmology, and therefore diminishing differences.    

Our box sizes are small in order to capture the clumpiness of the cold, pre-reionization gas, at the cost of missing large-scale power. We correct for this using the DC mode approach of \cite{Gnedin2011}, which allows us to model the effects of large-scale power.  We refer the reader to \cite{DAloisio2020} for a detailed description of our approach. In summary, we ran additional simulations with positive and negative box-scale overdensities applied to the cosmic mean density.  Quantities of interest (e.g. the LyC opacity) were obtained by integrating over the distribution of densities smoothed on the box scale.   We parameterize the box-scale over-density with $\delta/\sigma$, the linearly extrapolated density contrast in units of its standard deviation, smoothed on 2 $h^{-1}$Mpc scales.  In addition to our standard cosmic mean runs ($\delta/\sigma = 0$), we ran simulations with $\delta/\sigma = \pm \sqrt{3}$.  We get the mean LyC opacity by averaging over the distribution of densities from a cosmological N-body simulation (described in \S~\ref{subsec:neutral_islands}), assuming the opacity follows a power law in density between and outside our simulated values (see description of Eq.~\ref{eq:kappa_density}).\footnote{These values were originally chosen to apply the method of Gauss-Hermite Quadrature for the integration over the Gaussian distribution of linearly-extrapolated densities (see Appendix B of Ref \cite{DAloisio2020}).  }

\begin{figure*}
    \centering
   \resizebox{15cm}{!}{\includegraphics[trim={0cm 0cm 0cm 0cm},clip]{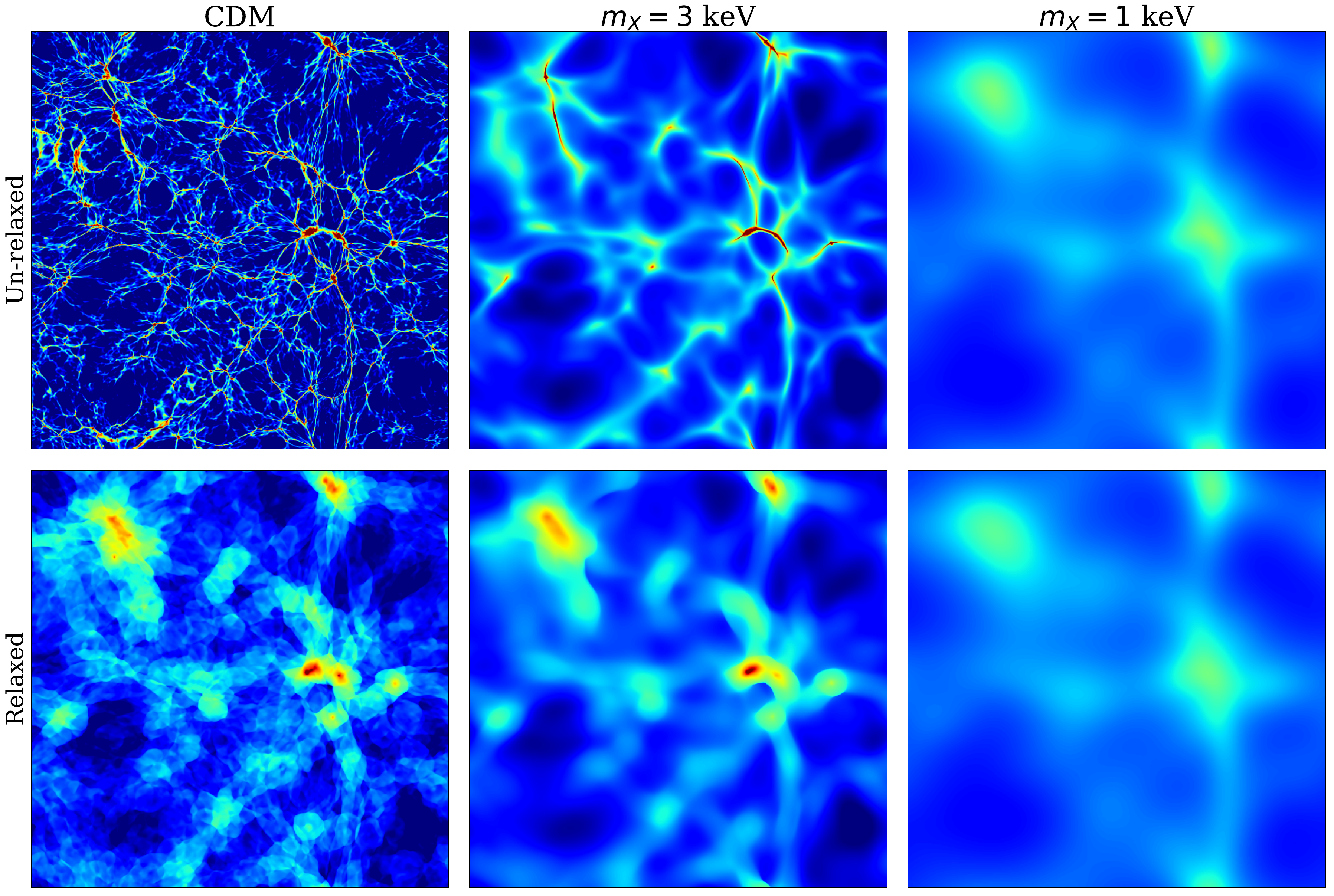}}
    \caption{
    Visualization of the density field at $z = 6$ for CDM (left), and WDM with $m_X = 3$ keV (middle) and $m_X = 1$ keV (right).  We show the un-relaxed runs in the top row and the relaxed runs in the bottom row.  In the un-relaxed limit, the CDM case differs dramatically from the WDM runs, with the gas clumping down to much smaller scales, resulting in a shorter mean free path.  However, in the relaxed limit, pressure smoothing and photoevaporation has mostly erased this extra structure, such that the mean free paths in the CDM and $m_X = 3$ keV density fields are similar.  In the WDM models, especially the $m_X = 1$ keV case, the density fields evolve considerably less than in CDM because small structures are largely missing to being with.  The lack of this small-scale structure in the $m_X = 1$ keV simulation results in a longer mean free path at all times.    
    }
    \label{fig:relax_vis}
\end{figure*}

Figure~\ref{fig:relax_vis} shows slices through the gas density field from our mean-density ($\delta/\sigma=0$) simulations at $z = 6$.  The top and bottom rows show the un-relaxed and relaxed limits respectively.  The columns show, from left to right, CDM, WDM with $m_X$ = 3 keV, and with $m_X$ = 1 keV.  Note the vast differences between CDM and the WDM cosmologies in the un-relaxed (cold, pre-reionization) limit; CDM initially clumps down to much smaller scales.  Comparing now the bottom panels, relaxation smooths the gas in CDM to a state not so dissimilar to that seen in the WDM run with $m_X$ = 3 keV. Note also that the WDM structures evolve less during relaxation because they lack the small-scale power from the start.  This is especially evident in the $m_X = 1$ keV run, for which the un-relaxed and relaxed limits are nearly indistinguishable by eye.

\subsection{Modeling the LyC opacity in ionized regions}
\label{subsec:reion_model}

We model the LyC opacity of the reionizing IGM as arising from two contributions: (1) Ionized gas within H II regions, including the self-shielding structures such as minihalos; (2) The neutral IGM that has yet to be reionized.  At $z\lesssim 6$, when the global neutral fraction is $\lesssim 10 \%$, these last remaining neutral regions are mainly relegated to structures of size $R\sim 10h^{-1}$Mpc that we will call ``neutral islands'' (see \S \ref{subsec:neutral_islands}).  The current section describes our procedure for modeling the opacity in ionized regions.   

We calculate the MFP at $912\text{\AA}$ in ionized gas directly from our simulations using the definition employed in \cite{Chardin2015}, 
\begin{equation}
    \label{eq:chardin_mfp}
    \mfp = -\left \langle \frac{\int x df}{\int df} \right \rangle = -\left \langle \int_{1}^{0} x df \right \rangle
\end{equation}
where $x$ is the distance along a sightline and $f = \exp(-\tau(x))$ is the factor by which ionizing photon flux would be attenuated along the sightline.  The second equality assumes that $f$ is negligible at the end of the sightline.  We evaluate Eq.~\ref{eq:chardin_mfp} directly by computing the integral for 10,000 randomly positioned and oriented sight lines and averaging the results.\footnote{We have checked that (1) the MFP as given by Eq.~\ref{eq:chardin_mfp} agrees well with the definition used in Ref \cite{DAloisio2020} and (2) 10,000 sightlines is sufficient for convergence of Eq.~\ref{eq:chardin_mfp}.  }  The gas fields in our un-relaxed runs are cold and fully neutral (because no ionizing background was applied).  To obtain the MFP in the limit of short $\Delta t$ after I-front passage, we post-processed the un-relaxed runs under the assumption of photoionization equilibrium assuming the case A recombination rate, applying also the self-shielding model discussed in the last section.  Applying the equilibrium assumption is motivated by the short photoionization time scale of $t_{\rm PI} \sim 1/\Gamma_{\rm HI}  \sim 100,000$ yr, relative to the tens to hundreds of Myr over which the relaxation process occurs.  We also set the temperature of the gas to a uniform $T_{\rm re}=20,000$ K, which is representative of temperatures in the wake of recently passed I-fronts \cite{1994MNRAS.266..343M, DAloisio2019, 2021ApJ...906..124Z}.  

\begin{figure}
    \centering
     \resizebox{12cm}{!}{\includegraphics[trim={0cm 0cm 0cm 0cm},clip]{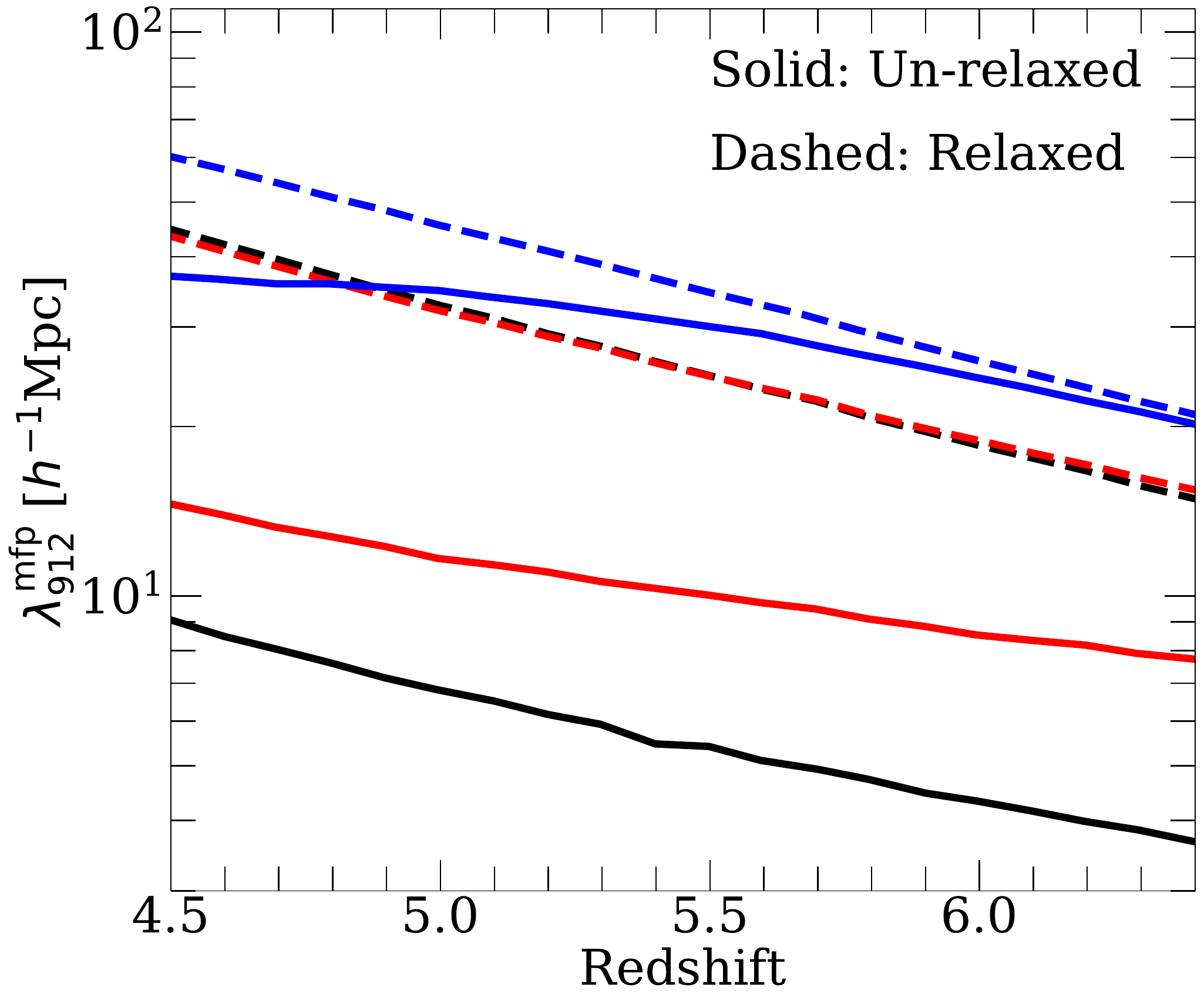}}
    \caption{Mean free path for a simple test case in which we post-processed our mean-density runs assuming a constant $\Gamma_{-12} = 0.3$ and $T = 10^4$ K for CDM (black), $m_X = 3$ keV (red) and $m_X = 1$ keV (blue), in the un-relaxed (solid) and relaxed (dashed) limits.   Holding $\Gamma_{\rm HI}$ and $T$ constant ensures that the evolution in the MFP reflects only changes in the density field.  There are considerable differences between the three DM models in the un-relaxed limit owing to the dramatic difference in the amount of small-scale structure shown in Figure~\ref{fig:relax_vis}.  However in the relaxed limit the CDM and $m_X = 3$ keV cases are nearly identical and the relative difference with the $m_X = 1$ keV run decreases, again reflecting the trends in Figure~\ref{fig:relax_vis}.  The similarity of the relaxed CDM and $m_X = 3$ keV cases highlights the fact that free-streaming and pressure smoothing affect the same mass scales in these models.  }
    \label{fig:relax_mfp}
\end{figure}

It is instructive the compare MFPs among our CDM and WDM simulations.  In Figure~\ref{fig:relax_mfp} we show such a comparison using the same runs from Figure~\ref{fig:relax_vis}.  To isolate differences arising from the different DM cosmologies (i.e. the gas density structures), we re-scale the photoionization rate in all of the simulations to a constant $\Gamma_{-12} = 0.3$, and we set the temperature to a uniform value of $T = 10^4$ K. In this case, evolution in the MFP is driven entirely by the density field.  The solid and dashed curves show the un-relaxed and relaxed limits, respectively. In the un-relaxed limit, the MFP increases substantially as the WDM particle mass decreases and the free-streaming scale increases. The relaxed limits are much more similar, with the CDM and $m_X = 3$ keV models being nearly identical at all redshifts.  This similarity between CDM and the $m_X = 3$ keV case follows intuitively from the similarity in the density strucure of the IGM seen in the bottom-left and bottom-middle panels of Fig. \ref{fig:relax_vis}.  The comparison also highlights that photoevaporation and pressure smoothing affect structures of the same mass scales that drive differences in the un-relaxed CDM and $m_X = 3$ keV runs.

Reionization is spatially patchy such that, at any given time during the process, different locations within the ionized phase of the IGM are at different stages of relaxation.  We will denote the redshift at which some patch of the IGM was reionized with $z_{\rm re}$.  The results of Figure \ref{fig:relax_mfp} suggest that the {\it local} MFP in a recently reionized patch ($z\sim z_{\rm re}$) of the IGM differs substantially between the DM models considered here.  However, the differences begin to disappear as the gas relaxes and small-scale power is erased.  Clearly, the sensitivity of the global MFP to the free streaming scale depends on the fraction of IGM that is still relaxing ($\Delta t \lesssim 300$ Myr since ionization).  The larger this fraction is, the more sensitive the global MFP will be to differences in small-scale power.  

Modeling the global mean free path at a given $z$ requires averaging over the distribution of local reionization redshifts, $z_{\rm re}$, in the ionized phase of the IGM.   Given a global reionization history, $x_{\rm ion}(z)$, this distribution at redshift $z$ can be written as

\begin{equation}
    \frac{dP}{dz_{\rm re}}(z, z_{\rm re}) = \frac{1}{x_{\rm ion}(z)}  \frac{dx_{\rm ion}}{dz_{\rm re}} \Big|_{z_{\rm re} \geq z}
    \label{eq:P(zre)}
\end{equation}
Our simulations provide models for the opacity of the ionized IGM in the un-relaxed and relaxed limits.  Denoting the local absorption coefficient at some location in the IGM as $\kappa(\Delta t,z_{\rm re})$, where $\Delta t$ is the cosmic time that has elapsed since $z_{\rm re}$, we model the evolution between the two limits with a simple relaxation ansatz, 

\begin{equation}
    \label{eq:kappaev}
    \kappa(\Delta t, z_{\rm re}) = \kappa_{\rm u} + [\kappa_{\rm r} - \kappa_{\rm u}]\left[1 - \exp\left(-\frac{\Delta t}{t_{\rm relax}}\right)\right].
\end{equation}
Here, $\kappa_{\rm u}$ and $\kappa_{\rm r}$ denote the un-relaxed and relaxed absorption coefficients, respectively, which are taken from our simulation runs (computed using Eq.~\ref{eq:chardin_mfp} with $\kappa \equiv 1/\mfp$), and $t_{\rm relax}$ is the relaxation time scale.  Note that $\kappa_{\rm u}$ and $\kappa_{\rm r}$ are the opacities averaged over simulation box-scale densities (DC modes), given by
\begin{equation}
    \label{eq:kappa_density}
    \kappa_{\rm X}(\Delta t, z_{\rm re}) = \int \kappa_{\rm X}(\Delta t, z_{\rm re}, \Delta_{\rm box})\frac{dP}{d\Delta_{\rm box}} d\Delta_{\rm box}
\end{equation}
where $X \in \{{\rm u},\rm{r}\}$, $\Delta_{\rm box}$ is the matter density smoothed on the box scale $L = 2$ $h^{-1}$Mpc, $\kappa_{\rm X}(\Delta t, z_{\rm re},\Delta_{\rm box}) = 1/\mfp$ is the absorption coefficient in the simulation box with box-scale density $\Delta_{\rm box}$, and $\frac{dP}{d\Delta_{\rm box}}$ is the PDF of $\Delta_{\rm box}$, which is obtained from the cosmological N-body simulation described in the next section.  
We interpolated $\log{\kappa}$ linearly in $\log{\Delta_{\rm box}}$, effectively assuming a power law relation between $\kappa$ and $\Delta_{\rm box}$.  \footnote{To convert between $\delta/\sigma$ and the nonlinear density $\Delta_{\rm box}$, we use Eq. 18 of Ref \cite{Mo1996}.  }

In what follows, we adopt a fiducial value of $t_{\rm relax} =150$ Myr unless otherwise noted.  This choice is consistent with the radiative hydrodynamics simulations of \cite{DAloisio2020}, in which relaxation is observed to be completed by $\Delta t \sim 300$ Myr.\footnote{Since $t_{\rm relax}$ is an e-folding timescale, relaxation will be $\sim 90$\% complete after 2$t_{\rm relax} = 300$ Myr in our fiducial model.  }  Note that $t_{\rm relax}$ essentially sets the relative importance of un-relaxed gas in our model.  Larger values result in more un-relaxed gas contributing to the opacity, which, in turn, amplifies differences between the DM cosmologies.  We will explore how different choices of $t_{\rm relax}$ affect our main results in \S \ref{subsec:astro}.  In Appendix~\ref{app:patchy}, we test our relaxation ansatz against the opacity evolution in a simulation run with $z_{\rm re} = 6.5$.  We find 10\% or better agreement with the simulation at $4.5 < z < 6.5$, confirming both the accuracy of the ansatz and our fiducial choice of $t_{\rm relax}$.

Finally, the average absorption coefficient in ionized gas is 

\begin{equation}
    \label{eq:avgkappaion}
    \langle \kappa_{\rm ion} \rangle (z) = \int_{z_{\rm init}}^{z} dz_{\rm re} \kappa(z, z_{\rm re}) \frac{dP}{dz_{\rm re}}(z, z_{\rm re}),
\end{equation}
where $z_{\rm init}$ is the starting point of reionization, which we take to be $z=12$.\footnote{Note that specifying $z$ and $z_{\rm re}$ is equivalent to specifying $\Delta t$ and $z_{\rm re}$.}  We note that the $\kappa(z, z_{\rm re})$ appearing in (\ref{eq:avgkappaion}) is already averaged over DC modes (Eq.~\ref{eq:kappa_density}), hence Equation~\ref{eq:avgkappaion} neglects the correlation between density and $z_{\rm re}$ due to inside-out reionization.  This may mean that we over-estimate the impact of un-relaxed gas, since at the end of reionization recently ionized gas is expected to be under-dense on average.  However, this effect is degenerate with uncertainties in $t_{\rm relax}$, since both affect the relative importance of un-relaxed gas, and hence should not impact our broad conclusions.  We also note that our analysis does not account for spatial fluctuations in $\Gamma_{\rm HI}$ in ionized gas, which may persist on large scales near the end of and after reionization~\cite{Davies2016,DAloisio2018}.  This effect is also likely degenerate with $t_{\rm relax}$ to some extent, since un-relaxed gas in voids is likely to also have the lowest $\Gamma_{\rm HI}$.   The absorption coefficient in ionized gas is added to a contribution coming from neutral islands, which we discuss in the next section. The global reionization histories from which $\frac{dP}{dz_{\rm re}}$ are obtained (see Eq. \ref{eq:P(zre)}) come from radiative transfer simulations of reionization, which we also describe in the next section.  

\subsection{Opacity from Neutral Islands}
\label{subsec:neutral_islands}

\begin{figure*}
    \centering
     \centering
     \resizebox{15cm}{!}{\includegraphics[trim={0cm 0cm 0cm 0cm},clip]{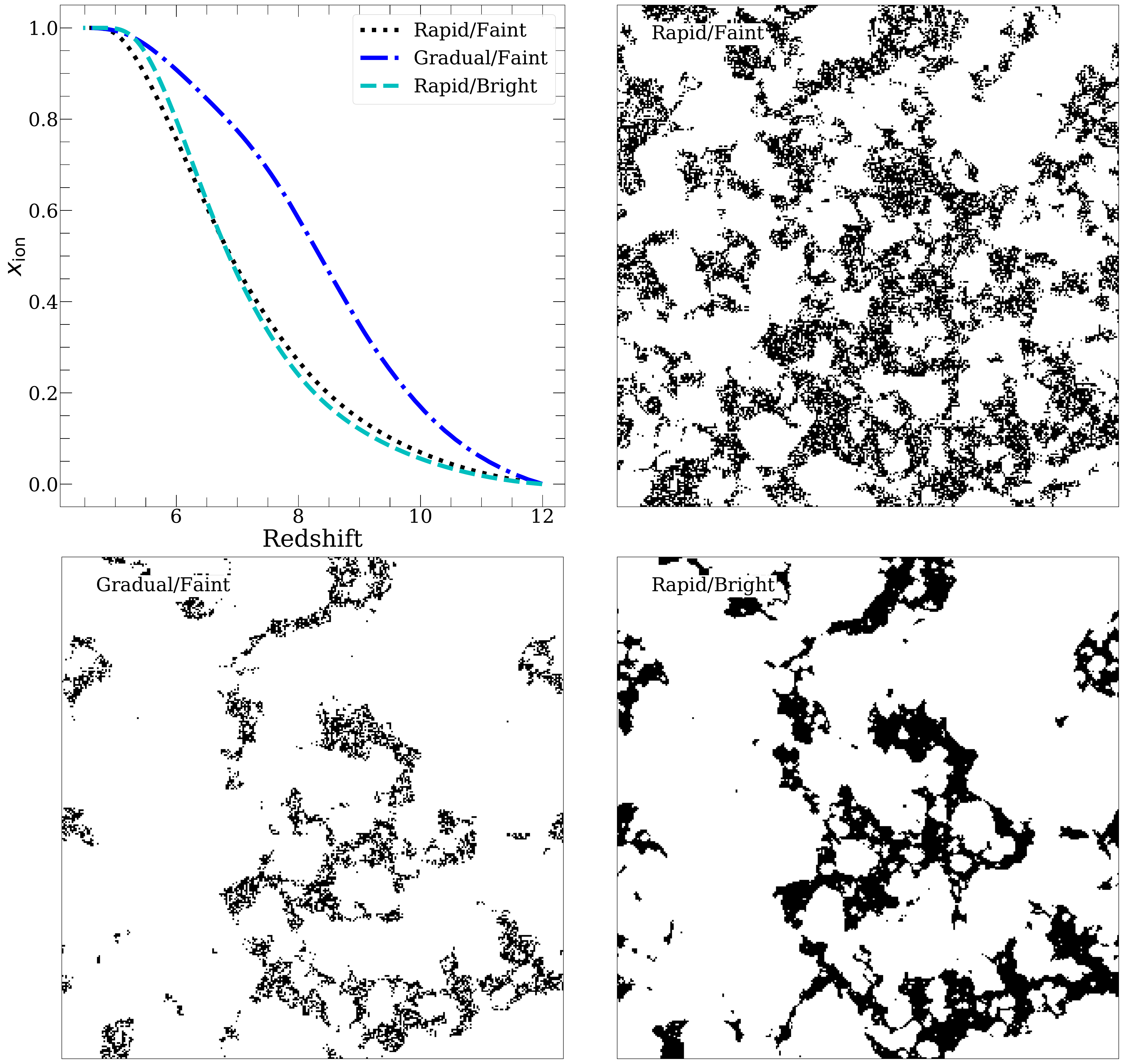}}
    \caption{Reionization simulations employed in this study.  {\bf Upper left}: volume-weighted mean ionized fraction vs. redshift for the three reionization histories considered in this work.  {\bf Other panels}: slices through the ionization fields at $z = 6$ in a ($300$ $h^{-1}$Mpc)$^3$ volume for each scenario.  Black denotes cells with HI fractions $\geq 0.5$.  We consider rapid reionization models in which reionization is driven by faint and bright sources (upper right and lower right respectively) and a gradual model in which reionization is nearly over at $z = 6$ (lower left).  The neutral islands are the smallest in the gradual case because the neutral fraction is smallest.  In the case where bright sources dominate, the neutral island are less porous and hence take up a smaller effective volume (at approximately fixed global neutral fraction). These simulations were run with the radiative transfer code of \cite{Cain2021}.  We use them to model the distribution of reionization redshifts, $\frac{dP}{dz_{\rm re}}$ (see eqs. \ref{eq:P(zre)} and \ref{eq:avgkappaion}), as well as the contribution to the LyC opacity from neutral islands. 
    }
    \label{fig:ion_histories}
\end{figure*}

We modeled the opacity from neutral islands using ionization fields from simulations of reionization.  These were run with the radiative transfer code of \cite{Cain2021} in a box with $L= 300 h^{-1}$Mpc and $N = 300^3$ RT cells.  The halo catalogs and density fields for the RT simulation were obtained from a cosmological DM-only simulation run with MP-Gadget~\cite{Feng2018}, with $N=2048^3$ DM particles.  Halos were identified on-the-fly with a friends-of-friends algorithm down to a minimum mass of $M^{\rm halo}_{\min} = 8.5\times10^9$ $h^{-1}$M$_{\odot}$, which corresponds to 32 DM particles.\footnote{We checked that the halo mass function in our simulation agrees well with published mass functions down to this limit.}  Models suggest that halos below this mass could have hosted galaxies that contributed significantly to reionization. Thus, we extended the RT source halos down to a minimum mass of $10^{9}h^{-1}M_{\odot}$ using a sub-grid algorithm based on the non-linear biasing approach of~\cite{Ahn2015} (see also \cite{Cain2021}).\footnote{To populate the sub-grid halos, we drew from the mass function of~\cite{Watson2013}, which agrees well with the resolved mass function in our simulation.} 

Several recent works have studied how the reionization source population differs in WDM cosmologies, compared to CDM \cite[e.g.][]{Villanueva-Domingo2018, Carucci2019, Romanello2021}.   In models with a larger free-streaming scale, reionization tends to start later and is driven by brighter sources owing to the suppression of the halo mass function at lower masses. In what follows, we do not model the effects of the underlying DM cosmology on the reionization history, with the rationale that any differences in the ionizing emissivity of the sources are mostly degenerate with uncertain astrophysical parameters such as the star formation efficiency or escape fraction.  We assume that we could always tune these source parameters to achieve approximately the same global neutral fraction among the DM cosmologies. Our main aim here is to quantify how small-scale power in the ionized IGM changes the mean free path, so we will compare different cosmologies {\it at a fixed global neutral fraction}. We note that our approach neglects potential differences in the neutral island morphology among the cosmologies.   

To explore how different reionization histories/morphologies come into play with the observed mean free path, we employed three models for the ionizing photon output of the sources.  The first is the fiducial rapid reionization model from~\cite{Cain2021}, in which reionization has a late midpoint ($z \sim 7.1$) and in which every halo down to with a minimum mass of $M_{\min}^{\rm halo} = 10^9$ $h^{-1}M_{\odot}$ has the same ionizing photon emissivity.  The latter condition means that the faintest, lowest-mass halos produce the bulk of the ionizing photons.  We refer to this scenario as the ``Rapid/Faint'' model.  Our second model has a more gradual reionization history with an earlier midpoint ($z \sim 8.5$), and also assigns the same emissivity to every halo - we refer to this as the ``Gradual/Faint'' model.  Our third model has a similar reionization history as the Rapid/Faint case, but has a minimum mass of $8.5 \times 10^9$ $h^{-1}M_{\odot}$ and assumes the emissivity of each halo is proportional to its UV luminosity, obtained by abundance matching to the UV luminosity function of~\cite{Finkelstein2019}.  In this case, reionization is dominated by bright, highly biased sources - we refer to this as the ``Rapid/Bright'' model.  What is important for our purposes is that the structure of the neutral regions in these models are significantly different, as shown visually in Figure \ref{fig:ion_histories}.  The top-left panel shows the global reionization histories in the three models, while the other panels show slices through the ionization fields at $z=6$.  Neutral gas is depicted in black. Comparing the top- and bottom-right panels, the two rapid models have approximately the same global neutral fraction of $x_{\rm HI} \approx 25\%$ at $z=6$. The model with brighter sources (bottom), however, exhibits less porous neutral islands and larger ionized regions compared to the model with fainter sources (top).    The neutral islands are much smaller in the gradual model because reionization is closer to completion, with $x_{\rm HI}(z=6) \approx 10\%$.   

\begin{figure}
    \centering
    \includegraphics[scale=0.20]{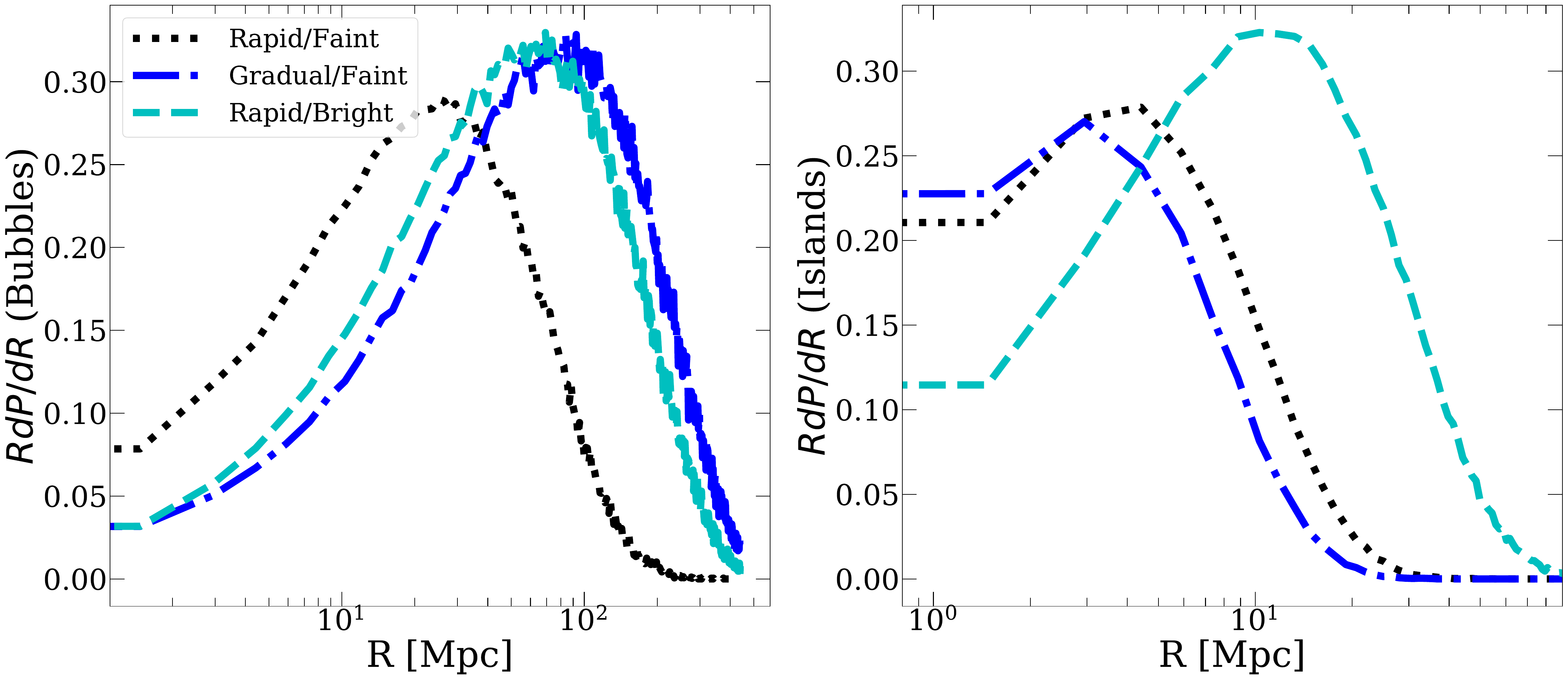}
    \caption{Distribution of ionized bubble (left) and neutral island (right) sizes in our reionization simulations at $z = 6$.  Both use the ray tracing method of~\cite{Mesinger2007} for quantifying the region size.  The Rapid/Faint case has the smallest ionized bubbles and hence the highest opacity due to neutral islands, while the other two models have similar bubble sizes.  The Rapid/Bright model has the largest (least fragmented) neutral islands owing to the sparsity and bias of sources in that model, while the other two models have smaller islands sizes.  }
    \label{fig:bubble_island}
\end{figure}

In Figure~\ref{fig:bubble_island}, we show the ionized bubble and neutral island size distributions (IBSD and NISD, left and right panels, respectively) in our simulations at $z = 6$.  We define these sizes using the ray-tracing method described in Ref~\cite{Mesinger2007} and implemented in the publicly available package {\it tools21cm}~\cite{Giri2018}.  The Rapid/Faint model has the smallest ionized bubbles, as is clearly seen in Fig.~\ref{fig:ion_histories}, and thus the highest opacity due to neutral gas.  In the right panel the Rapid/Bright model stands out with the largest (least fragmented) neutral islands owing to the sparse and highly biased distribution of its ionizing sources.  Note that the x axis is in Mpc here rather than $h^{-1}$Mpc to aid comparison with the NISD results of Refs~\cite{Giri2019b,Wu2021b}.\footnote{On average our islands are somewhat smaller than found by~\cite{Giri2019b} at $25\%$ and $10\%$ neutral (left and middle panels of their Fig. 9) while our Rapid/Bright model is similar to the models in~\cite{Wu2021b} at $16\%$ neutral (left panel of their Fig. 6).  Our islands may be smaller than theirs in part because of our threshold of $x_{\rm HI} > 0.5$ for a cell to be part of an island, which ignores many partially ionized cells, especially in the models with faint sources.  Still, the spread {\it between} our models is similar to the range found in~\cite{Giri2019b} and larger than that in~\cite{Wu2021b}, giving us confidence that our range of scenarios is representative.}   WDM scenarios with lighter $m_X$ have fewer low-mass halos (see left panel of Fig.~\ref{fig:mfp_iso} in \S\ref{sec:enhancedpk}) and thus we expect them to have morphologies more toward our Rapid/Bright case.  As we will discuss below, the degree to which neutral islands affect the measured MFP is uncertain, potentially in a way degenerate with morphology differences between WDM models.  Given these uncertainties, for simplicity we proceed by comparing different cosmologies with the neutral island morphologies fixed.  

We used the 3-dimensional ionization fields to calculate the contribution from neutral islands to the mean absorption coefficient. Following \cite{Cain2021}, we traced 50,000 sight lines from random positions to create mock quasar absorption spectra, and then extracted the absorption coefficient by fitting a stack of these spectra to the model of~\cite{Prochaska2009}. This stacking/fitting procedure mimics the method by which the mean free path is measured observationally.  When we fit the stacked absorption spectra following~\cite{Prochaska2009}, we allow the normalization of the flux to float in the fit, as is done in~\cite{Becker2021}.  Importantly, this prevents sightlines that start in neutral islands from contributing to $\langle \kappa_{\rm neutral} \rangle$, since they contribute zero flux to the stack at all wavelengths.  In contrast, fixing the normalization of the stack to be $1$ at $912 \text{\AA}$ effectively includes these sight lines (inappropriately) and yields a somewhat higher $\langle \kappa_{\rm neutral} \rangle$ (lower MFP).  For the Rapid/Faint, Gradual/Faint, and Rapid/Bright models respectively, our fiducial procedure yields a neutral island MFP that is $43$\%, $16$\%, and $29$\% larger than obtained by fixing the stack normalizations to $1$.  To isolate the contribution from neutral islands, we set the opacity to zero in cells where the local ionized fraction is $x_{\rm ion} > 0.5$.  Denoting the absorption coefficient from neutral islands with $\langle \kappa_{\rm neutral}\rangle$, we write the total absorption as $ \kappa_{\rm global} = \langle \kappa_{\rm ion}\rangle+\langle \kappa_{\rm neutral}\rangle$, and then take the ``measured'' mean free path to be $\mfp = 1/\kappa_{\rm global}$.  We caution, however, that our estimate of $\mfp$ likely does not provide an exact apples-to-apples comparison to the measurements of Refs~\cite{Becker2021,Bosman2021b} at $z = 6$.  We will return to this point in the next section.  

\section{DM models with suppressed small-scale power (WDM)} \label{sec:results}

\subsection{Results}
\label{subsec:results}

\begin{figure}
    \centering
    \includegraphics[scale=0.22]{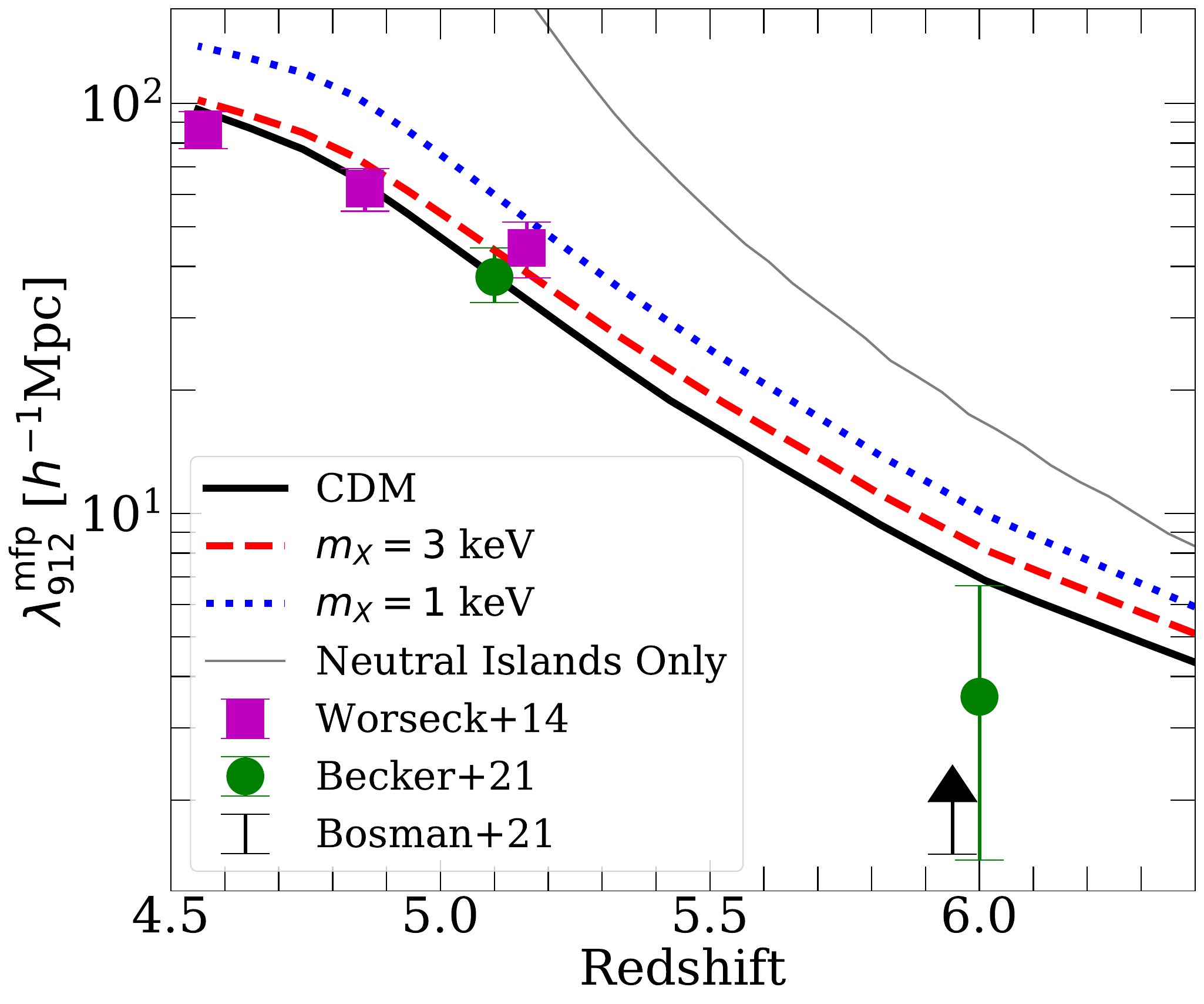}
    \includegraphics[scale=0.22]{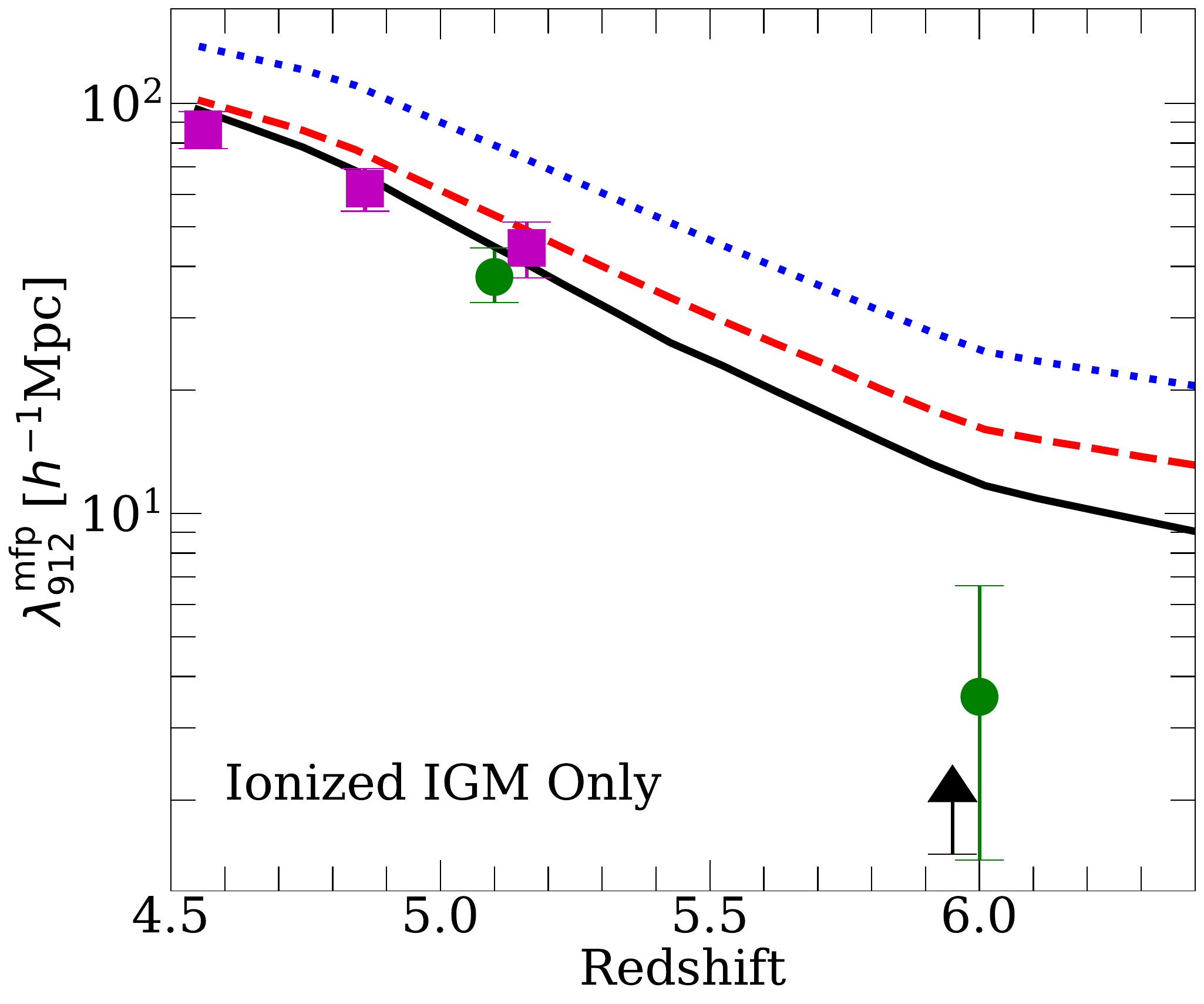}
    \caption{Mean free path for CDM and two WDM models with different particle masses.  
    {\bf Left}: Mean free path evolution for CDM (solid black), $m_X = 3$ keV (dashed red) and $m_X = 1$ keV (dotted blue) in our faint sources/rapid reionization scenario.  The thin grey line shows the MFP due to netural islands alone ($1/\langle \kappa_{\rm neutral} \rangle$, see \S \ref{subsec:neutral_islands}). The results shown here represent our full model of the IGM opacity, including contributions from ionized regions and neutral islands.  We show the observational measurements of Refs~\cite{Worseck2014,Becker2021,Bosman2021b} (the lower limit from Ref~\cite{Bosman2021b} at $z = 6$ has been shifted slightly to the left for clarity).  {\bf Right}: the MFP for the same three models but with the neutral island contribution omitted.  
    Differences between the DM cosmologies are suppressed by two effects: (1) at lower redshift ($z \lesssim 5.5$), Jeans pressure smoothing and photo-evaporation erases much of the small-scale power that would otherwise distinguish these models; (2) At higher redshift $z\gtrsim 6$, neutral islands contribute increasingly to the IGM opacity.  Hence, at fixed global neutral fraction, the MFPs become more similar between the models when the effects of neutral islands are included (compare left and right panels).  }
    \label{fig:mfp_mX}
\end{figure}

Figure~\ref{fig:mfp_mX} shows the redshift evolution of the MFP in the CDM and two WDM models considered here.  For this comparison, we use the Rapid/Faint reionization scenario described above, but we will explore the other scenarios below.  The data points show the observational measurements and limits of Refs \cite{Worseck2014}, \cite{Becker2021} and \cite{Bosman2021b}. The left panel shows our full models, including opacity from ionized gas and neutral islands, while the right panel considers only the opacity from ionized gas.   The thin gray curve in the left panel corresponds to the opacity from neutral islands, i.e. $1/\langle \kappa_{\rm neutral} \rangle$ as defined in \S\ref{subsec:neutral_islands}.   As expected, the MFP generally increases with the free streaming scale.  The differences are rather modest, however, in the left panel. At $z=6$, the model with $m_X = 3~(1)$ keV differs from the CDM case by $19$ ($45$) \%.  By $z=4.55$, the difference is $5$ ($43$) \%.

The differences between the DM cosmologies are smaller than we might naively expect, particularly at $z\gtrsim 6$, where the prevalence of un-relaxed/clumpy gas should drive larger differences between CDM and WDM.  To understand why, consider the MFP neglecting the contribution from neutral islands (right panel). There, the differences are indeed larger at higher redshift, when a larger fraction of the gas is un-relaxed. However, comparing to the left panel, the opacity from neutral islands contributes more at these redshifts. Most importantly, {\it at fixed global neutral fraction}, the neutral islands contribute a larger share of the opacity as the free streaming scale is increased.  Thus, when they are accounted for, the neutral islands obscure differences arising from the free streaming scale. This effect would be even larger if the free streaming scale also suppressed the source population.  For example, one could imagine a scenario in which reionization is delayed in the model with $m_X = 1$ keV.  In this case, the neutral islands would play an even larger role in setting the opacity at $z=6$, perhaps bringing $\mfp$ closer to the CDM result. 

On the other hand, by $z\sim 4.5$, most of the small-scale structure in the CDM model has been erased by smoothing/photo-evaporation.  We thus see the CDM and $m_x = 3$ keV models converging.   Visually, this is consistent with the bottom- left and -middle panels of Fig. \ref{fig:relax_vis}. We are led to conclude that the marked lack of difference seen in the left panel owes to two effects: (1) neutral island opacity at higher redshift ($z\gtrsim 6$); (2) relaxation at lower redshift ($z\lesssim 5$).   Both the CDM and $m_X = 3$ keV results agree reasonably well with MFP measurements at $z \leq 5$, while the $m_X = 1$ keV case overshoots by a factor of $\sim 1.5$. This suggests that it may be difficult to reconcile the $m_X = 1$ keV model with the measurements unless reionization ends even later than in our fiducial model, for which $x_{\rm ion}\approx 20\%$ at $z=6$.  

\begin{figure}
    \centering
    \includegraphics[scale=0.22]{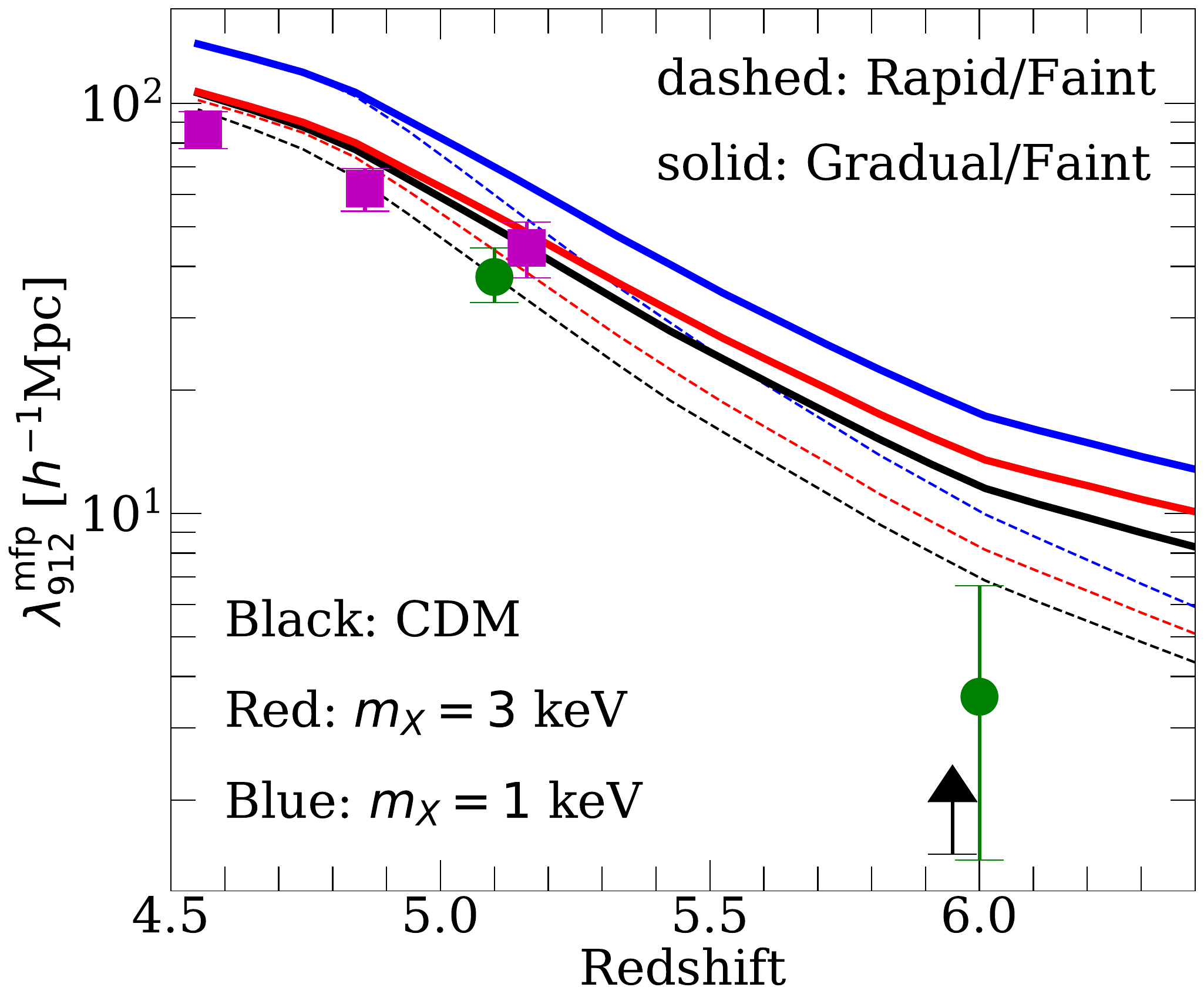}
     \includegraphics[scale=0.22]{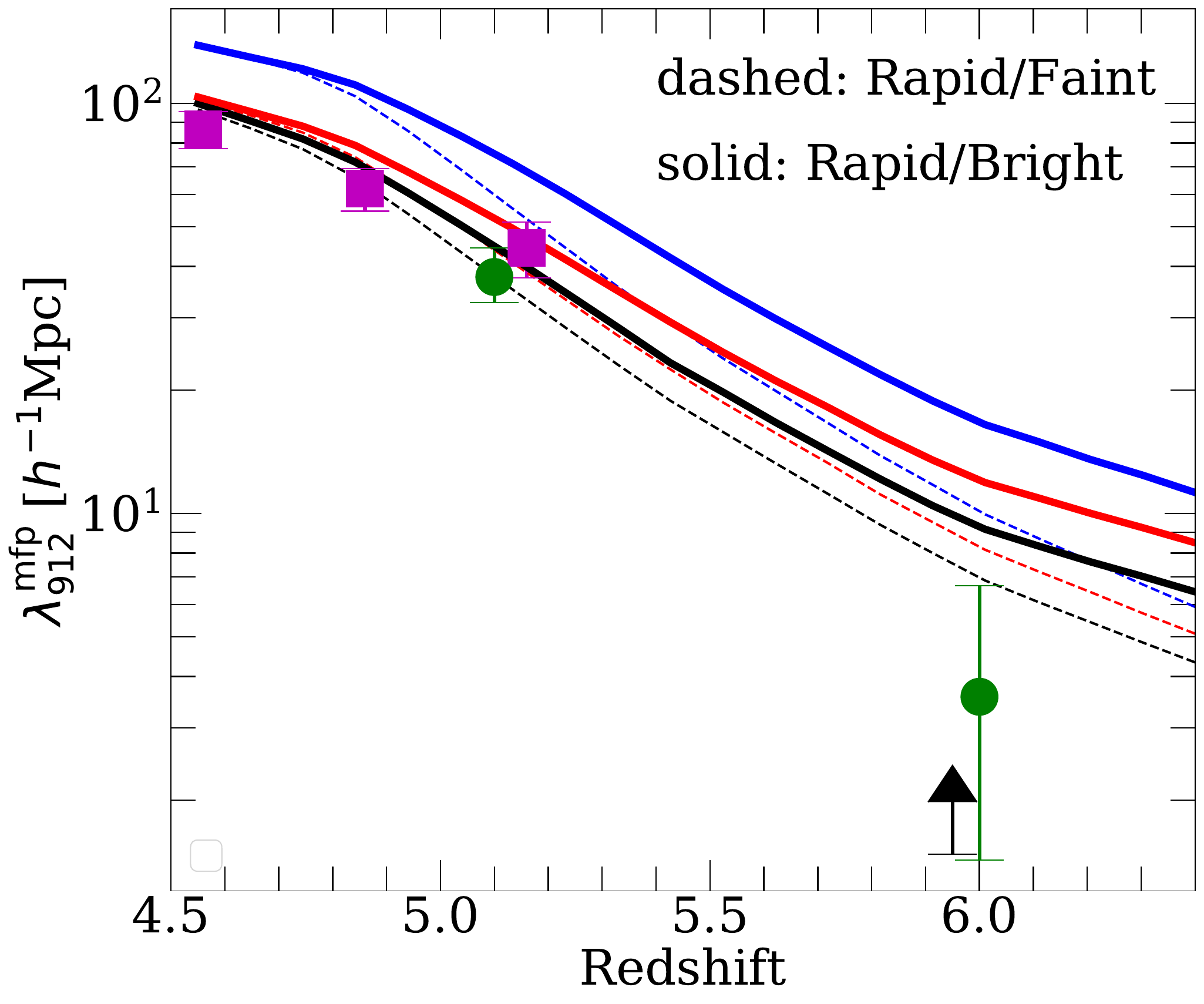}
    \caption{Interplay between the DM cosmology and features of the reionization model.
    {\bf Left}: Comparing our Gradual/Faint (solid) and Rapid/Faint (dashed) reionization scenarios.  Both models here assume our fiducial source model, which is driven by faint galaxies (see main text, and Fig. \ref{fig:ion_histories}). In the Gradual/Faint case the MFP is longer at $z = 6$ owing to the reduced opacity from neutral islands, and to the dearth of clumpy, un-relaxed gas in the ionized regions.  The latter also suppresses differences between the DM cosmologies.   {\bf Right}: Comparing our Rapid/Faint and Rapid/Bright scenarios.  These models have the same global reionization histories, but vary the brightness and bias of the sources.   The MFP is longer in the Rapid/Bright model because the neutral islands are more anti-biased, so they contribute less to the total IGM opacity. In this case, the relative differences between DM models are enhanced because the small-scale structure in ionized regions contributes a larger fraction of the total opacity.  
    }
    \label{fig:mfp_reion}
\end{figure}

Figure \ref{fig:mfp_reion} shows what happens if we vary the underlying reionization model.   The left panel compares our Gradual/Faint (solid) and Rapid/Faint (dashed) reionization scenarios. The global neutral fractions at $z=6$ are $x_{\rm HI}\approx$10\% and $20$\%, respectively.  Overall, the MFP at $z=6$ is significantly longer in the Gradual/Faint model for two reasons: (1) A smaller contribution from neutral islands owing to the smaller $x_{\rm HI}$; (2) A larger fraction of relaxed gas in the ionized regions, since much of the IGM is reionized earlier in the Gradual/Faint model (see top-left panel of Fig. \ref{fig:ion_histories}). Neutral islands contribute less to the opacity at $z=6$ in the gradual model, which would act to enhance differences between the WDM and CDM models.  This effect is muted, however, because the ionized gas is, on average, more relaxed.  Hence there is less small-scale structure to drive a difference between the CDM and WDM $\mfp$.      

The curves in the right panel correspond to our Rapid/Faint and Rapid/Bright scenarios.  These have nearly identical reionization histories, so there is no significant difference in the relaxation state of the gas.  Rather, the differences in the MFP at $z = 6$ are driven entirely by the structure of the neutral islands.  In the Rapid/Bright scenario, there are fewer neutral islands and they are larger, on average, resulting in a significantly lower opacity contribution from neutral islands.  Hence, the WDM results differ more from the CDM case at $z = 6$, with the MFP being $30$\% and $80$\% larger for $m_X = 3$ and $1$ keV, respectively.  
These results highlight a key point for interpreting the measurement of $\mfp(z=6)$ by \cite{Becker2021}. Constraining the global neutral fraction at $z=6$ is of utmost importance for gaining insight into the sinks from the MFP measurement.  

Indeed, it would be helpful to know whether {\it any} of the opacity comes from neutral islands.   As mentioned in the previous section, our estimate of $\mfp$ near $z = 6$ may not be directly comparable to the measurements of~\cite{Becker2021,Bosman2021b}, mainly because it is unclear to what extent neutral islands, if they are present at $z = 6$, affect those measurements.  Most of the quasar spectra used in these works do not show evidence of neutral islands near the quasar, as is to be expected for the highly biased regions in which these quasars likely reside~\cite{Mesinger2010}.  As such,  their measurements may more closely reflect $\mfp$ in the ionized component of the IGM (see right panel of Fig.~\ref{fig:mfp_mX}).   

\begin{table}
    \centering
    \begin{tabular}{|c||c|c|}
        \hline
        \% Diff. at $z = 6$ for $m_X = 3$ $(1)$ keV & Ionized Only & Fiducial Prescription\\
        \hline
        \hline
        Rapid/Faint & 37 (112)\% & 19 (45)\% \\
        Gradual/Faint & 22 (69)\% & 17 (50)\% \\
        Rapid/Bright & 40 (121)\% & 30 (80)\% \\
         \hline
    \end{tabular}
    \caption{Percentage differences between the $m_X = 3$ ($1$) scenarios and CDM at $z = 6$ for all three of our reionization histories under different assumptions about the contribution of neutral islands to the measured MFP.  We bracket the range of possibility by assuming that either neutral islands do not contribute at all to the measured opacity (``Ionized Only"), or that they contribute as predicted by our fiducial stacking/fitting procedure (``Fiducial Prescription").  Although the differences between DM cosmologies are larger in the Ionized Only scenario, the truth is likely somewhere between these two limiting cases.  }
    \label{tab:neutral_percentages}
\end{table}

Table~\ref{tab:neutral_percentages} brackets the range of possibilities for the effect of neutral islands on our results. The middle column shows percentage differences in the MFP between WDM models and CDM for the scenario in which neutral islands do not contribute at all to the opacity.  The right column shows the same but adopting our fiducial prescription for the neutral island opacity.   As mentioned previously, the differences between the cosmologies are larger if the quasar stacks effectively measure the opacity of only the reionized phase of the IGM. The truth is likely somewhere between these two cases. This discussion highlights the need for further work on how neutral islands affect the measured MFP during reionization.   

\subsection{Effects of modeling assumptions}
\label{subsec:astro}

\begin{figure}
    \centering
    \includegraphics[scale=0.22]{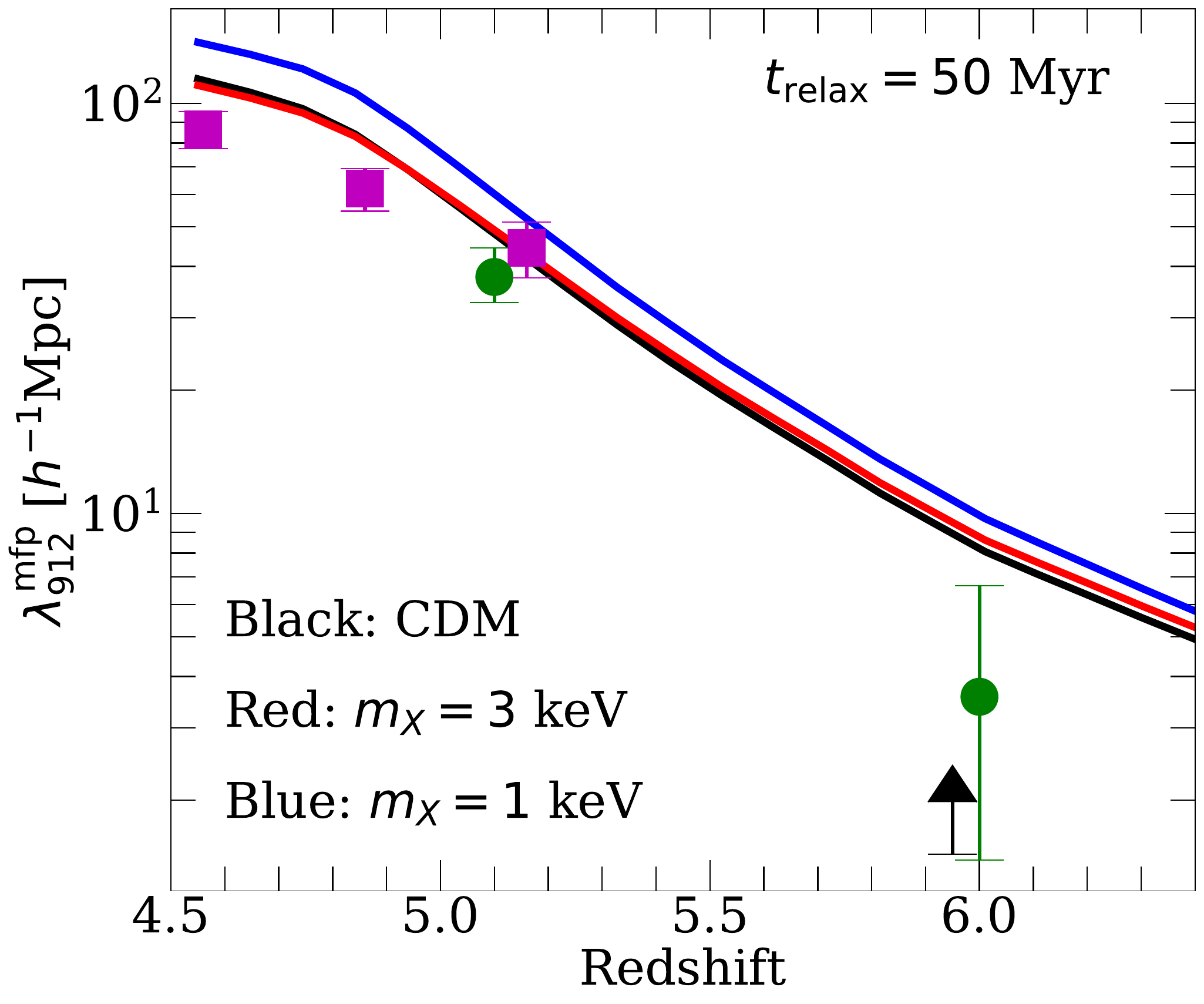}
    \includegraphics[scale=0.22]{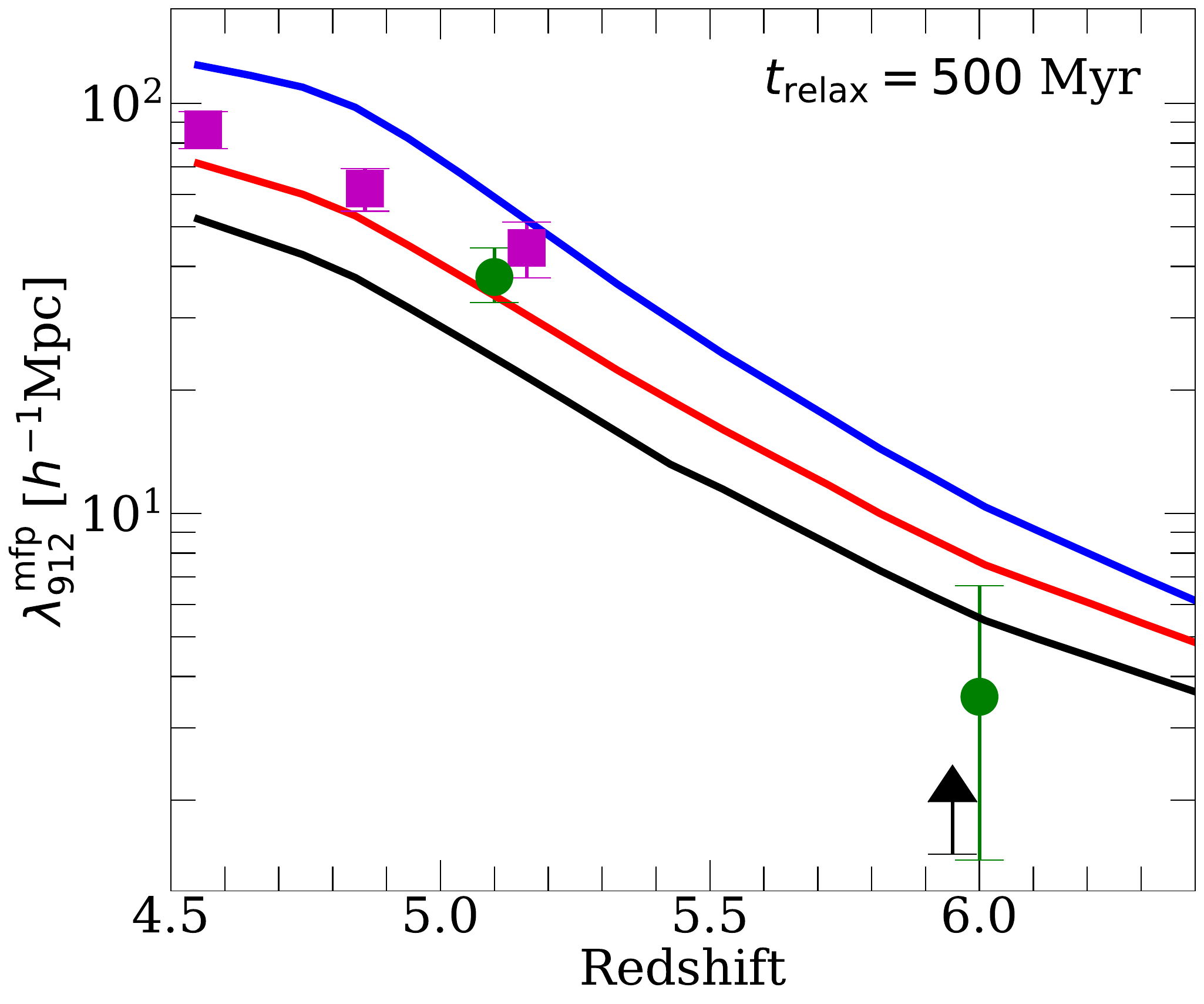}
    \caption{Effect of the relaxation timescale on the MFP.  {\bf Left}: MFP for each of our DM models assuming a relaxation timescale of $t_{\rm relax} = 50$ Myr.  {\bf Right}: same, but for $t_{\rm relax} = 500$ Myr.  We see that shorter (longer) relaxation timescales result in longer (shorter) $\mfp$ and a reduced (increased) difference between the different DM scenarios.  All these trends owe to the increased contribution to the opacity from un-relaxed gas in the scenario with larger $t_{\rm relax}$.  
    }
    \label{fig:mfp_trelax}
\end{figure}

In this section, we will examine the effects of several assumptions made in our modeling.   We adopt our Rapid/Faint model in the ensuing comparisons.  In Figure~\ref{fig:mfp_trelax}, we show the effect of varying the gas relaxation timescale in our model, $t_{\rm relax}$.  Recall that lengthening (shortening) this timescale enhances (reduces) the contribution of un-relaxed gas to the opacity, which acts to increase (decrease) differences in $\mfp$ between the WDM and CDM models.  The left panel assumes $t_{\rm relax} =50$ Myr, while the right panel assumes $t_{\rm relax} =500$ Myr.  These values were chosen to be somewhat extreme examples to highlight the effect of this parameter.  (Our fiducial value is $t_{\rm relax} =150$ Myr, which is motivated by recent results from radiative hydrodynamics simulations \cite{Park2016,DAloisio2020}.) We see that the difference between the DM models is generally greater when the relaxation time scale is longer, i.e. when a large fraction of the IGM is un-relaxed.  For example, relative to the CDM $\mfp$ at $z=6$, the models with $m_X = 3$ ($1$) keV  have a $37$ ($90$) \% longer $\mfp$ for $t_{\rm relax} = 500$ Myr.  For  $t_{\rm relax} = 50$ Myr, the difference reduces to $7$ ($20$) \%.  The CDM results with $t_{\rm relax} = 500$ Myr are more consistent with the short MFP at $z = 6$, but they under-shoot the $z \leq 5$ measurements.  We note, however, that a value of $t_{\rm relax} = 500$ Myr is a much longer time scale than is observed in the simulations of \cite{DAloisio2020}. Figure~\ref{fig:mfp_trelax} mainly highlights that the sensitivity of the MFP to small-scale power relies on how much of the ionized IGM is un-relaxed.

\begin{figure}
    \centering
    \includegraphics[scale=0.205]{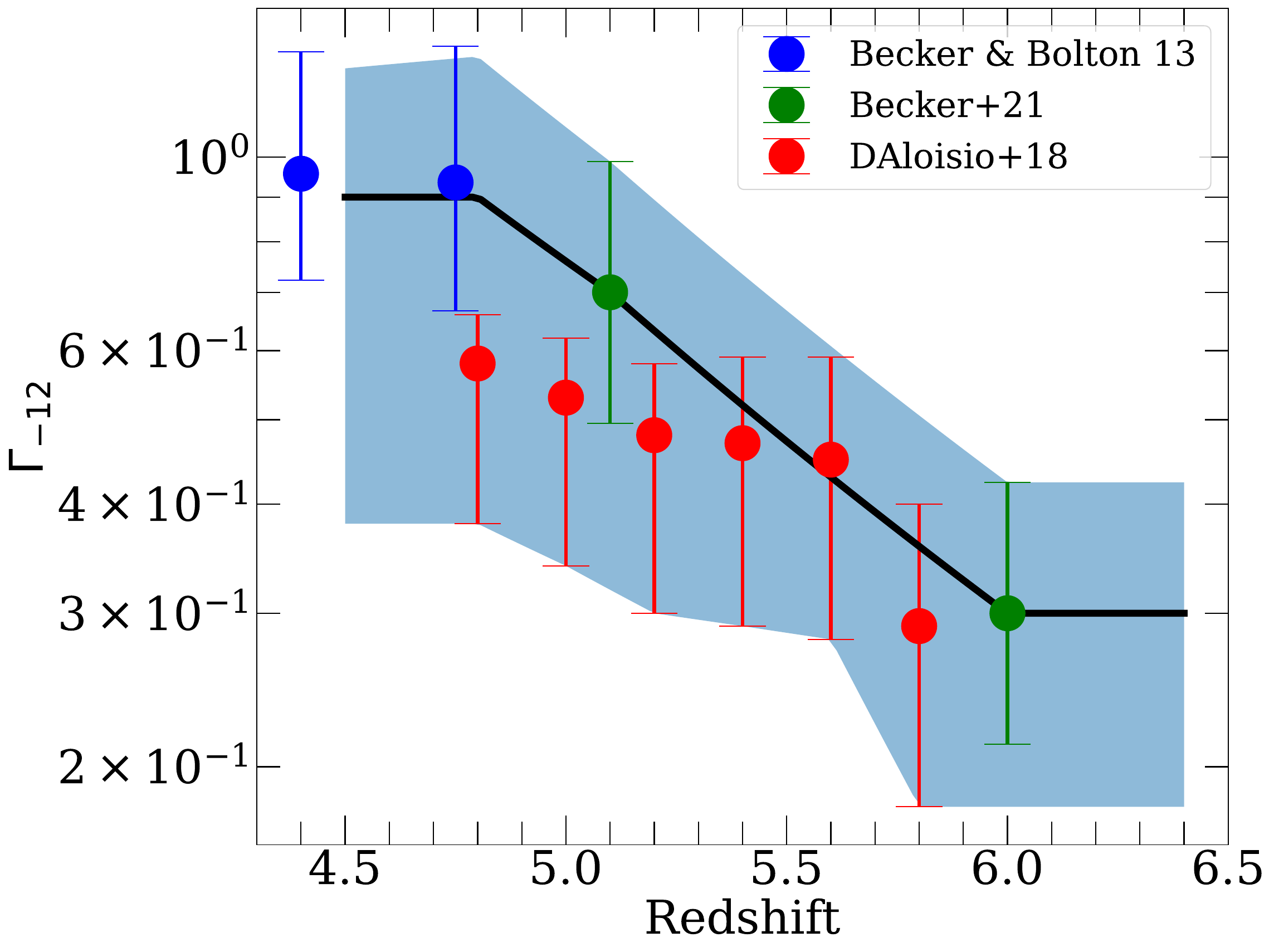}
    \includegraphics[scale=0.205]{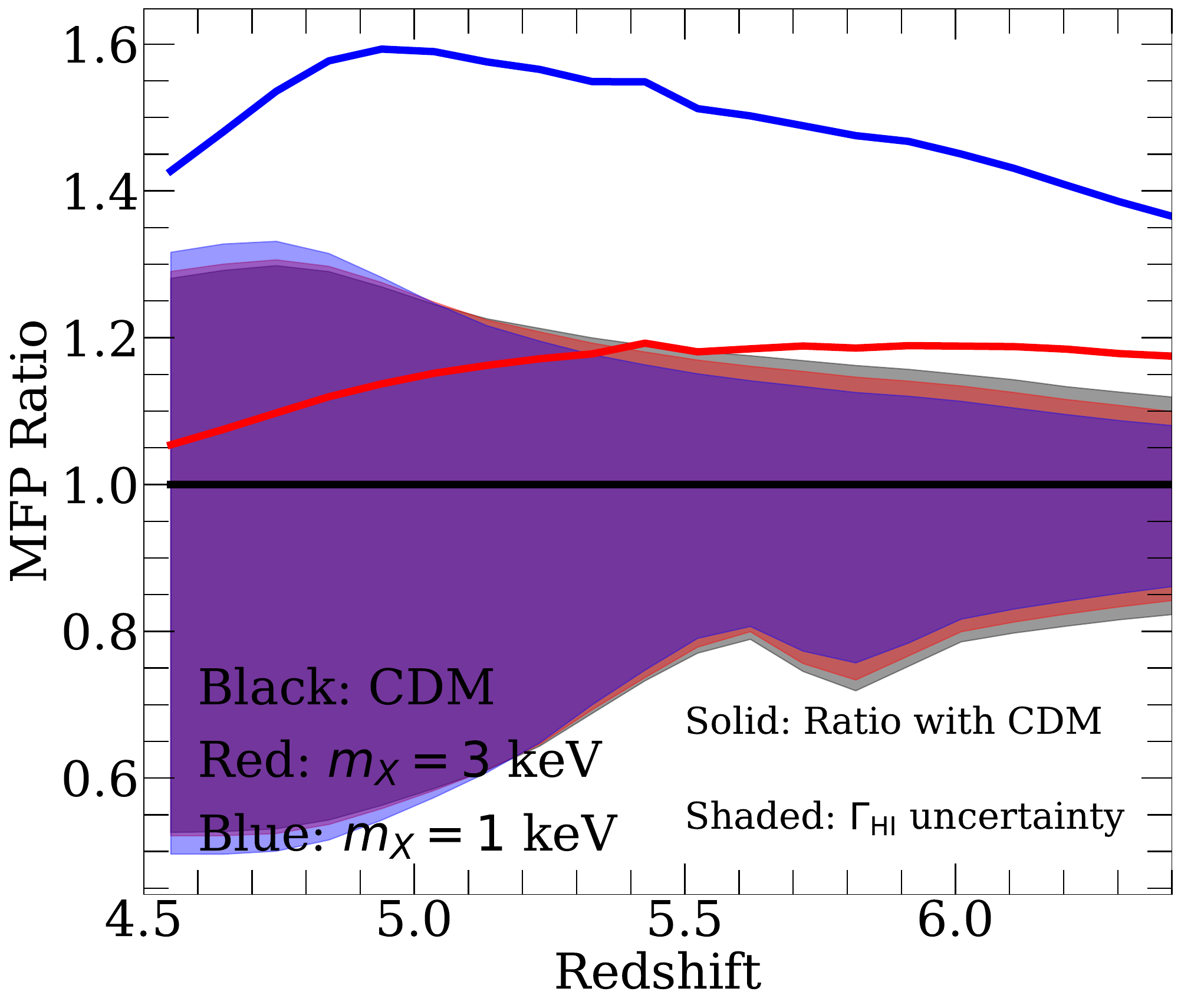}
    \caption{Effect of different $\Gamma_{\rm HI}$ histories on the MFP.  {\bf Left}: range of $\Gamma_{\rm HI}$ histories considered in our analysis (shaded region) compared to three sets of measurements from the literature.  The solid black line denotes our fiducial history for $\Gamma_{\rm HI}$.  {\bf Right}: Comparison of the uncertainty in the MFP from DM models vs. $\Gamma_{\rm HI}$.  The solid lines denote the ratio of the MFP with CDM for the different DM models assuming the fiducial $\Gamma_{\rm HI}$ history, and the shaded regions denote the range spanned by the different $\Gamma_{\rm HI}$ histories for each model.  At $z > 5$ the size of the spread due to $\Gamma_{\rm HI}$ is similar to the difference between CDM and $m_X = 1$ keV, and at $z < 5$ the former is significantly larger.  The shaded regions all roughly overlap, highlighting the insensitivity of the $\Gamma_{\rm HI}$ uncertainty to the assumed DM cosmology.  We see that uncertainty in the history of $\Gamma_{\rm HI}$ would considerably complicate any effort to distinguish even the CDM and $m_X = 1$ keV scenarios, the latter of which has been ruled out already by the Ly$\alpha$ forest. }  
    \label{fig:mfp_gamma}
\end{figure}

Next we consider how the intensity of the extragalactic ionizing background factors into our calculations. The data points in the left panel of Figure \ref{fig:mfp_gamma} show Ly$\alpha$ forest measurements of the hydrogen photoionization rate, which scales with the intensity of the ionizing background.   Clearly there is still considerable uncertainty in $\Gamma_{-12}$ at $z \gtrsim 4.5$.  The solid/black curve shows the evolution of $\Gamma_{-12}$ in our hydrodynamic simulations of the sinks. Here we explore what happens to our results if we vary $\Gamma_{-12}$.  We bracketed the uncertainties in the measurements with two histories in $\Gamma_{-12}$, shown as the upper and lower bounds of the blue shading in Figure \ref{fig:mfp_gamma}.  We then re-scaled the neutral hydrogen densities in our simulations under the assumption of photoionization equilibrium, and recomputed the MFPs in our models.  The right panel of Figure \ref{fig:mfp_gamma} shows the result of this exercise.  For each DM model, the shaded region corresponds to the ratio of the MFP from the high and low $\Gamma_{\rm HI}$ histories with the fiducial one.  The solid curves denote the ratio of the MFP with CDM for each of the DM models assuming the fiducial $\Gamma_{\rm HI}$ history (that is, of the solid curves in Fig.~\ref{fig:mfp_mX}).  The range spanned by the shaded regions is similar to the difference between CDM and $m_X = 1$ keV at $z > 5.5$ (a factor of $\sim 1.4$), and becomes significantly larger at $z < 5.5$ (a factor of $\sim 2.5$ vs. $\sim 1.5$).  These results suggest that the considerable uncertainties in $\Gamma_{-12}$ alone make it difficult to rule out convincingly with MFP measurements even the $m_X = 1$ keV model, which has already been ruled out by Ly$\alpha$ forest flux power spectrum measurements \cite{Irsic2017}. 

Lastly, we comment on the assumed thermal history of the IGM in our models.  Consider first the thermal history of a gas parcel starting with the impulsive heating by I-fronts at redshift $z_{\rm re}$ (we will discuss the history prior to this below).  At $z \leq z_{\rm re}$, the thermal history of the gas affects the evolution of its Jeans filtering scale as well as its equilibrium \HI\ fraction through the temperature dependence of the recombination rate.  The simple relaxation ansatz employed here (eq. \ref{eq:kappaev}) implicitly models these effects by evolving the opacity in hot, un-relaxed gas (assumed to be at $T_{\rm re} = 20,000$ K), towards a cooler, relaxed limit with $T\sim 8,000$ K, the temperature to which the gas relaxes in our hydrodynamic simulations. Variation in the thermal history at $z \leq z_{\rm re}$ can come in two forms: (1) variation in the impulsive heat injection by I-fronts as they sweep through, encapsulated in the reionization temperature, $T_{\rm re}$; (2) variation in the photoionization equilibrium heating rate of the highly ionized gas, well after I-front passage.  The latter depends only on the spectrum of the extragalactic ionizing background, which maintains the ionization state of the gas.  While this spectrum is highly uncertain, a realistic variation in the spectral shape leads to only $\sim 30 \%$ changes in the IGM temperature \cite{McQuinn2016, DAloisio2019}, and a smaller change in the opacity/MFP.  Variations in $T_{\rm re}$ could lead to factor of $2$ changes in the thermal history if reionization is still ongoing at $z < 6$ (see e.g. Fig. 1 of \cite{DAloisio2019}).  Using radiative transfer simulations of I-fronts, Ref \cite{DAloisio2019} found that reionization temperatures are likely in the range $T_{\rm re} \approx 17,000 - 30,000$ K.   We note that the upper limit of this range,  $T_{\rm re} = 30,000$ K, is $50\%$ hotter than our assumed value. Since the amount of heat dumped into the IGM at $z_{\rm re}$ is $\sim k_B T_{\rm re}$, we expect no more than a $\sim 50\%$ effect on our results for a realistic variation $T_{\rm re}$.  Note also that variations at this level are much smaller than the factor of $20 - 1,000$ jump in temperature that occurs when an I-front sweeps through a cold, neutral region.

Variation in the thermal history of the gas {\it before} $z_{\rm re}$, however, could have a larger effect on our results because it alters the Jeans filtering scale of the un-relaxed gas, which is the driver of differences in $\mfp$ between CDM and WDM. Our hydrodynamic simulations assume that the gas was cooling adiabatically after kinematic decoupling from the CMB. As a result, the gas can get as cold as $\sim 10$ K before the impulsive heating to $T_{\rm re}$.  If, for example, the first X-ray sources were extremely efficient at pre-heating the gas, then the IGM could have started out with far less small-scale structure on $\sim 10^4 - 10^6$ M$_\odot$ scales, perhaps lengthening the predicted MFP at $z=6$.  To quantify roughly the maximum effect of this pre-heating on IGM clumping, Ref \cite{DAloisio2020} considered an extreme scenario among their radiative hydrodynamic simulations, in which the gas was not allowed to cool below $T=1,000$ K below $z=20$.  They found a suppression in the IGM clumping factor of $\sim 1.9$ within the first $\sim 10$ Myr of $z_{\rm re}$, which translates to a lengthening of $\mfp$ by the same factor.   Although this assessment likely overestimates the effect considerably,   the thermal history of the gas prior to reionization is an important uncertainty further obscuring the cosmological interpretation of high-$z$ MFP measurements. 

We conclude this section by discussing briefly how uncertainties from modeling assumptions affect the feasibility of constraining WDM with MFP measurements.  In a fully ionized IGM, the main factor limiting constraints is $\Gamma_{\rm HI}$. Figure~\ref{fig:mfp_gamma} shows that given current measurements, $\Gamma_{\rm HI}$ produces factor of $\sim 2$\ uncertainty in the MFP at $z < 5.2$.  Under very optimistic assumptions assumptions about the longevity of un-relaxed gas (right panel of Figure~\ref{fig:mfp_trelax}, $t_{\rm relax} = 500$ Myr), the difference between CDM and $m_X = 3$ ($1$) keV is a factor of $\sim 1.4$ ($2$) shortly after reionization ends.  Assuming $\mfp \propto \Gamma_{\rm HI}^{2/3}$, which holds approximately in our calculations, the uncertainty in $\Gamma_{\rm HI}$ should be reduced by at least a factor of $\sim 1.7$ to produce a spread similar to the difference between CDM and $m_X = 3$ keV.  More realistic values of $t_{\rm relax}$ suggest that this factor would be even larger.\footnote{We note also that if the IGM was still undergoing reionization at $5 < z< 6$, it is unclear how to interpret existing measurements of $\Gamma_{\rm HI}$.  These measurements generally rely on models/simulations to map the measured forest transmission to $\Gamma_{\rm HI}$. The models employed to date do not include neutral islands and fluctuations from reionization (see however Ref \cite{DAloisio2018} for a discussion of the latter).}

At $z > 5.2$, there is only one measurement ($z = 6$), which has large error bars.  Even if future efforts can populate $5.2 < z < 6$ with measurements of similar fidelity to those at $z < 5.2$ (a challenging task), getting constraints would be complicated by uncertainties in the reionization history and morphology.  
As Figure~\ref{fig:mfp_reion} illustrates, the possible scenarios are sufficiently diverse that even if the detailed redshift evolution of the MFP were known, breaking the degeneracy with a poorly constrained reionization history could be challenging.  So to obtain constraints from this range of redshifts, even with high-quality MFP measurements, some constraints on the global neutral fraction and a better understanding of how neutral islands contribute to the {\it measured} LyC opacity would be required.

In summary, the predicted difference in $\mfp$ between CDM and WDM cosmologies depends critically on how much gas is in an un-relaxed (still clumpy) state, which, in turn, depends on the uncertain reionization history and time scale for relaxation.   Furthermore, large uncertainties in the intensity of the extragalactic ionizing background, and in the thermal history of the gas prior to reionization, would further weaken cosmological inferences from high-$z$ MFP measurements. Based on these considerations, we conclude that the observed short value of $\mfp(z=6)$ is unlikely in the near future to be a useful no-go test of DM models with small-scale power cutoffs.  

\section{Models with enhanced small-scale power}
\label{sec:enhancedpk}

Some alternatives to standard CDM, such as ultra-light axions~\citep{Irsic2019, Xiao2021}, or primordial black holes \cite{Clesse2017}, predict a shot noise-like enhancement in the small-scale matter power spectrum.  This class of scenarios is of interest for the current paper owing to the shortness of the measured $\mfp(z=6)$.    Indeed, the measurements of \cite{Becker2021} are pushing models toward a very late and rapid reionization process, and it is unclear at present whether this picture can be reconciled in a physically consistent way with the evolution of the Ly$\alpha$ forest flux evolution and its spatial fluctuations \cite{Cain2021, Garaldi2021, Lewis2022}.  This motivates exploring the role that additional small-scale power could have in producing short values of $\mfp$, a task that we take up here.  

\begin{figure}
    \centering
    \includegraphics[scale=0.21]{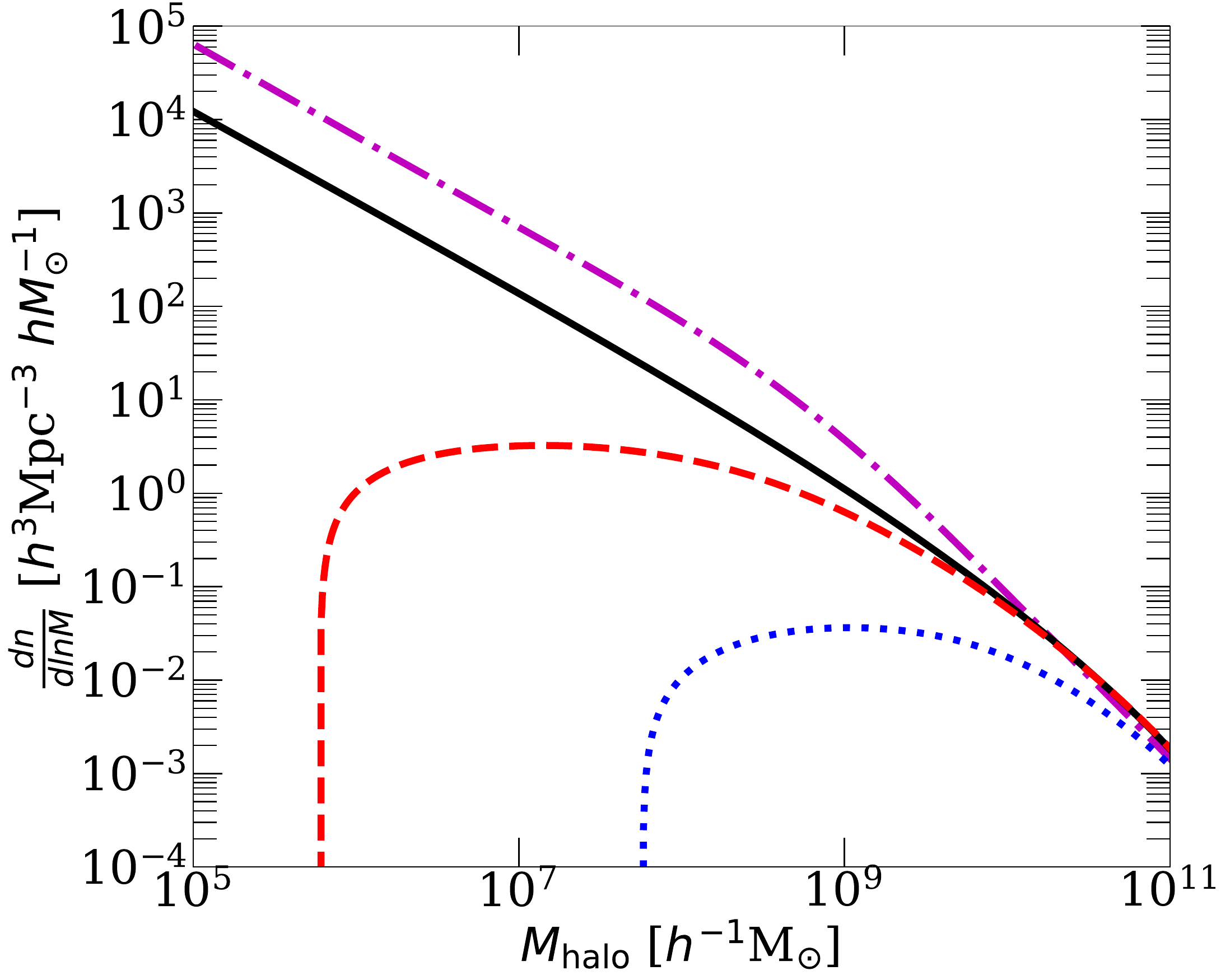}
    \includegraphics[scale=0.21]{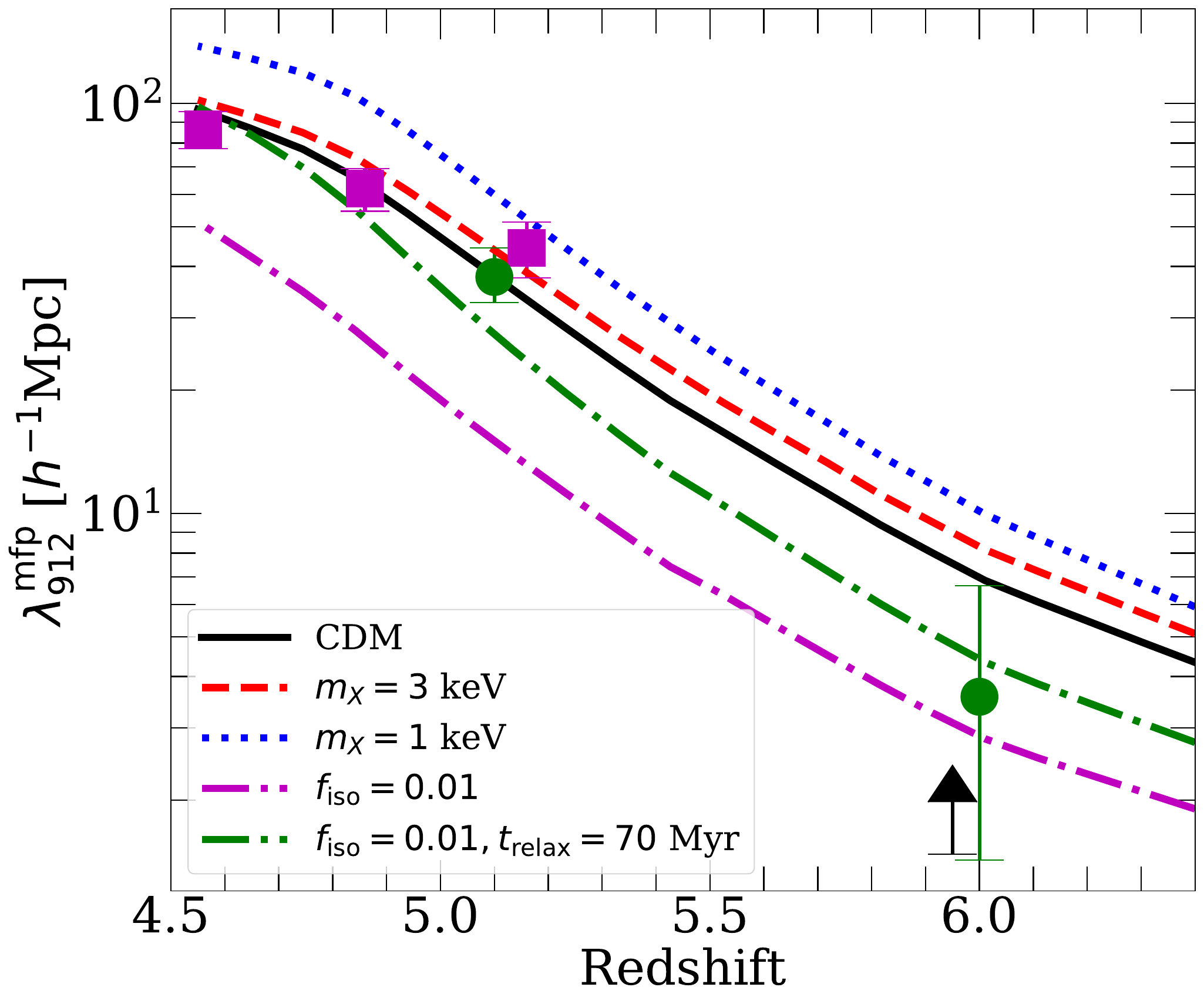}
    \caption{
    MFP evolution in DM cosmologies with enhanced small-scale power. As an illustrative model, we use the axion-like scenario considered in \cite{Irsic2019}.  {\bf Left}: halo mass function for the CDM case (black solid curve) compared against the axion-like scenario (dot-dashed magenta) and the thermal relic WDM models with $m_X = 3$ and $1$ keV (dashed red and dotted blue).  {\bf Right}: the MFP for the axion-like case compared to the other DM models assuming $M_{\min} = 10^4$ $h^{-1}M_{\odot}$ in equation \ref{eq:kappahalo}, all assuming the Rapid/Faint reionization scenario.  The enhanced power scenario has a factor of $\sim 7$ more halos at masses $\leq 10^8$ $h^{-1}M_{\odot}$ than the CDM case, which contribute significant additional opacity in the un-relaxed limit.  The resulting MFP at $z = 6$ is in better agreement with the central value of the measurement, but under-shoots the measurements at lower redshifts considerably. This discrepancy can be ameliorated by modifying the reionization history or adopting a shorter relaxation timescale.  The green dot-dashed curve shows an example of the latter, with $t_{\rm relax} = 70 $ Myr (compared to our fiducial value of $t_{\rm relax} = 150$ Myr).   
    }
    \label{fig:mfp_iso}
\end{figure}

As a representative example, we use the ultra-light axion DM scenario considered in Ref~\cite{Irsic2019} in which there is a white noise contribution to the power spectrum from isocurvature fluctuations.  We adopt their $f_{\rm iso} = 0.01$ model, which corresponds to a significant enhancement in halo abundance on $\lesssim 10^8 M_{\odot}$ scales, the mass range of interest for this study.  The orange curve in Figure~\ref{fig:powerspec} shows the linearly extrapolated power spectrum in this model. In contrast to the previous section, we did not run hydrodynamic simulations in the ultra-light axion cosmology.  Instead, we use a simplistic model to estimate the effect of the enhanced power on the opacity of un-relaxed gas. In the picture adopted here, the halos are treated as dense, optically thick ``billiard ball'' absorbers, with aggregate absorption coefficient 
\begin{equation}
    \label{eq:kappahalo}
    \kappa_{\rm halo} = \int_{M_{\min}}^{\infty} dM \sigma_{h}(M) \frac{dn}{dM}.
\end{equation}
Here, $dn/dM$ is the halo mass function and $\sigma_h(M)$ is the physical cross-section of a halo with mass $M$, which we approximate to be
\begin{equation}
    \label{eq:sigmam}
    \sigma_{\rm h}(M) = \pi R_{\rm 200}^2(M),
\end{equation}
where $R_{\rm 200}(M) = \left[ 3 M / (4 \pi \times 200 \rho_c(z)) \right]^{1/3}$ is the halo virial radius and $\rho_{\rm c}(z)$ is the critical density.  The parameter $M_{\rm min}$ is the mass of the smallest {\it gaseous} halo to form, which roughly corresponds to the Jeans filtering scale of the gas. In the un-relaxed limit, we take $M_{\rm min} = 10^4~h^{-1}$M$_\odot$, the Jeans scale of the adiabatic hydrodynamic simulations that we use to model the un-relaxed IGM.   To evaluate equation \ref{eq:kappahalo} we use the halo mass function from Ref \cite{Tinker2008}, which is calibrated using the same definition of the virial radius ($R_{200}$) that we employ here.  The left panel of Figure \ref{fig:mfp_iso} compares the mass function in the axion-like model to CDM and the two WDM models of the last section.  The axion-like model exhibits enhanced halo formation on small-scales, but is nearly indistinguishable from CDM above $\sim10^{10}$ M$_\odot$.  % $\sim10^9$ M$_\odot$.  

If we assume that the halos provide all the opacity to ionizing photons, the relative difference between the CDM and axion-like scenarios can be evaluated directly from the above expressions.  This assumption is closest to true in the un-relaxed limit where most of the opacity is sourced by dense, self-shielding absorption systems in the mass range $\sim 10^4 - 10^8$ M$_{\odot}$~\citep{Nasir2021}.  As more time elapses since $z_{\rm re}$, photoevaporation and relaxation of the small-scale power will drive the $\mfp$ in the two models closer, as we have already seen in the previous section.  We proceed here by assuming that the ratio $\kappa_{\rm halo}^{f_{\rm iso} = 0.01}/\kappa_{\rm halo}^{\rm CDM}$ captures the enhancement over CDM in the un-relaxed limit, and that the relaxed limit of the two models are the same.  To implement this in our relaxation model, we simply re-scale the $\kappa_u$ for CDM in equation (\ref{eq:kappaev}) by the ratio $\kappa_{\rm halo}^{f_{\rm iso} = 0.01}/\kappa_{\rm halo}^{\rm CDM}$, while leaving $\kappa_r$ the same as before.  
The results of this procedure are shown as the magenta dot-dashed curve in the right panel of Figure~\ref{fig:mfp_iso}, alongside our CDM and WDM models, all assuming the Rapid/Faint reionization scenario. The green dot-dashed curve corresponds to the same $f_{\rm iso} = 0.01$ model, but adopting a shorter relaxation timescale of $t_{\rm relax}= 70$ Myr.  Raising $M_{\min}$ to a value of $10^6$ $h^{-1}$M$_{\odot}$ changes $\kappa_{\rm halo}^{f_{\rm iso} = 0.01}/\kappa_{\rm halo}^{\rm CDM}$ very little across all redshifts, although the individual values of $\kappa_{\rm halo}$ change by nearly an order of magnitude. 

The enhanced small-scale power in the axion-like model produces significantly better agreement with the central value of the $z = 6$ MFP for our fiducial $\Gamma_{\rm HI}$ and reionization histories.   However, at least for our fiducial choice of $t_{\rm relax} = 150$ Myr, the agreement at $z \lesssim 5$ is considerably worse than for CDM and WDM with $m_X = 3$ keV.  The reason is that the extremely clumpy un-relaxed gas in the axion-like model that brings the MFP down at $z = 6$ stays un-relaxed for too long and drives the MFP too low at $z \leq 5$.  This effect can be ameliorated to some degree by modifying the reionization history to give the gas more time to relax and/or by invoking a shorter relaxation timescale.  The latter is illustrated by the green dot-dashed curve, with $t_{\rm relax} = 70$ Myr.   In the axion-like model, the relaxation timescale could in fact be shorter than our fiducial choice of $t_{\rm relax} = 150$ Myr (which was motivated by radiative hydrodynamic simulations in a CDM cosmology), if the opacity is dominated by very small clumps/halos that are quickly photo-evaporated.    

We conclude by noting that better agreement with the $z=6$ MFP could also be achieved within CDM given the large uncertainties in astrophysical parameters discussed previously.  For example, if $\Gamma_{\rm HI}(z = 6)$ were a factor of $\sim 2$ lower than our fiducial value (see e.g. Ref~\cite{Lewis2022}) our CDM prediction would fall closer to the measured central value of the MFP, without compromising the agreement at lower redshift.  Given this consideration as well as large uncertainties in the MFP measurements themselves, we caution against over-interpreting Figure~\ref{fig:mfp_iso} at this time.  These results mainly serve to quantify the potential role of cosmology in setting the intergalactic LyC opacity.

\section{Conclusion} 
\label{sec:conc}

Ref \cite{Becker2021} recently presented a measurement of the mean free path of ionizing photons at $z=6$.  At face value, the short value reported by them can be interpreted as evidence that the IGM clumps on scales $M \lesssim 10^8$ M$_\odot$, raising the possibility of using $\mfp(z=6)$ to rule out DM models lacking small-scale power.  Motivated by this, we have studied the role that the underlying DM cosmology plays in setting the $z> 5$ mean free path.   We considered thermal relic WDM as an example of models with a cutoff in small-scale power, and an ultralight axion candidate as an example with enhanced  power.  We compared these models against CDM predictions.  The main takeaways from this study can be summarized as follows:
 
\begin{itemize}
    \item   Many viable DM candidates exhibit stark differences with CDM on mass scales $10^4 \lesssim M/[M_{\odot}] \lesssim 10^8$. Gaseous halos in this range contribute much of the IGM opacity to ionizing photons immediately after a region has been reionized.  These structures are erased over a timescale $\sim 300$ Myr by photoevaporation and pressure smoothing. We therefore expect DM cosmologies to exhibit the largest differences in $\mfp$ during or shortly after reionization.  Recent models place the end of reionization around $z=5.2$, making $z>5$ $\mfp$ measurements a potential window into the underlying DM model.

    \item In our thermal relic WDM scenarios with particle mass $m_{\rm X} = 3~(1)$ keV, the $z=6$ $\mfp$ in {\it ionized gas} is $37~(112)~\%$ longer than in CDM.  However, at  fixed global neutral fraction, the contribution to $\mfp$ from neutral islands acts to obscure these differences.  For example, in our fiducial reionization model with neutral fraction $\approx 20\%$ at $z=6$, we found more modest differences of $19~(45)~\%$ in $\mfp$ when the contribution from neutral islands is included. Hence, without knowing the global neutral fraction, it is difficult to rule out DM models with a small-scale power cutoff. 
        
    \item Scenarios is which reionization ends earlier exhibit smaller differences in $\mfp$ between DM cosmologies. This owes to photoevaporation/pressure smoothing having more time to erase the small-scale structure that distinguishes these cosmologies.

    \item
    At fixed global neutral fraction, scenarios in which the neutral islands are larger, more clustered, and fewer in number lead to larger differences in $\mfp$ between DM cosmologies.  In these cases, the neutral islands contribute comparatively less to the IGM opacity, such that small-scale power in ionized regions plays a bigger role in setting $\mfp$.

    \item The enhanced small-scale power in the axion-like model lowers the predicted MFP at $z=6$, bringing it into better agreement with the central value measured by Ref~\cite{Becker2021} compared to CDM at fixed $\Gamma_{\rm HI}$ and global neutral fraction.

\item Among the key uncertainties precluding a robust conclusion on cosmology are the intensity of the extragalactic ionizing background and the thermal history of the IGM prior to reionization.  The former sets the densities which self-shield.  The latter sets the smallest gaseous structures that can form, the Jeans filtering scale. Variations in these quantities within plausible models produce differences in $\mfp$ similar to those observed among the DM cosmologies considered here.    
\end{itemize}

Our results illustrate the role that small-scale power plays in setting the MFP during reionization. A key consideration that arises from our analysis is the relative importance of neutral islands in setting $\mfp$ at $z=6$.   The less neutral islands contribute, the more opacity must come from small-scale power in ionized regions.  This question may be addressable with existing quasar absorption spectra and obviously has important implications for reionization itself. Another more basic question is whether our model of IGM opacity arising entirely from cosmological fluctuations is fundamentally correct.  If processes related to high-$z$ galaxy formation affect the physical state of intergalactic gas at large, then our models may be missing important physics shaping the sinks.  This question can be imminently addressed with more detailed hydrodynamic simulations.

\appendix

\section{Self-Shielding Implementation}
\label{app:selfshielding}

We account for self-shielding by using the results of~\cite{Nasir2021} to model $\Gamma_{\rm HI}(n_{\rm H})$ at high densities.  We find that the form $\frac{\Gamma_{\rm HI}}{\langle \Gamma_{\rm HI} \rangle} = F(n_{\rm H})\exp[-(n_{\rm H}/n_0)^6]$
where $F(n_{\rm H})$ is given by~\citep[as in][]{Rahmati2013, Chardin2018}
\begin{equation}
    \label{eq:selfshield}
    F(n_{\rm H}) = (1-f)\left(1 + \left[\frac{n_{\rm H}}{n_0}\right]^\beta\right)^{\alpha_1} + f\left(1 + \frac{n_{\rm H}}{n_0}\right)^{\alpha_2}
\end{equation}
fits well the median $\Gamma_{\rm HI}(n_{\rm HI})$ of Ref~\cite{Nasir2021} in relaxed gas for $(n_0/{\rm cm}^{-3}, \beta, \alpha_1, \alpha_2, f) = (0.015, \\2, -3, -1, 0.01)$ for $\Gamma_{-12} = 0.3$.  We additionally found that their result for $\Gamma_{-12} = 3.0$ can be reproduced by assuming the same parameters with $n_0 \propto \Gamma_{\rm HI}^{2/3}$, so we adopted this scaling to account for evolution of $\Gamma_{\rm HI}$.  We used the same self-shielding prescription in the post-processing MFP calculation for both the relaxed and un-relaxed simulations.  

\section{Numerical Convergence}
\label{app:convergence}

\begin{figure}
    \centering
    \includegraphics[scale=0.16]{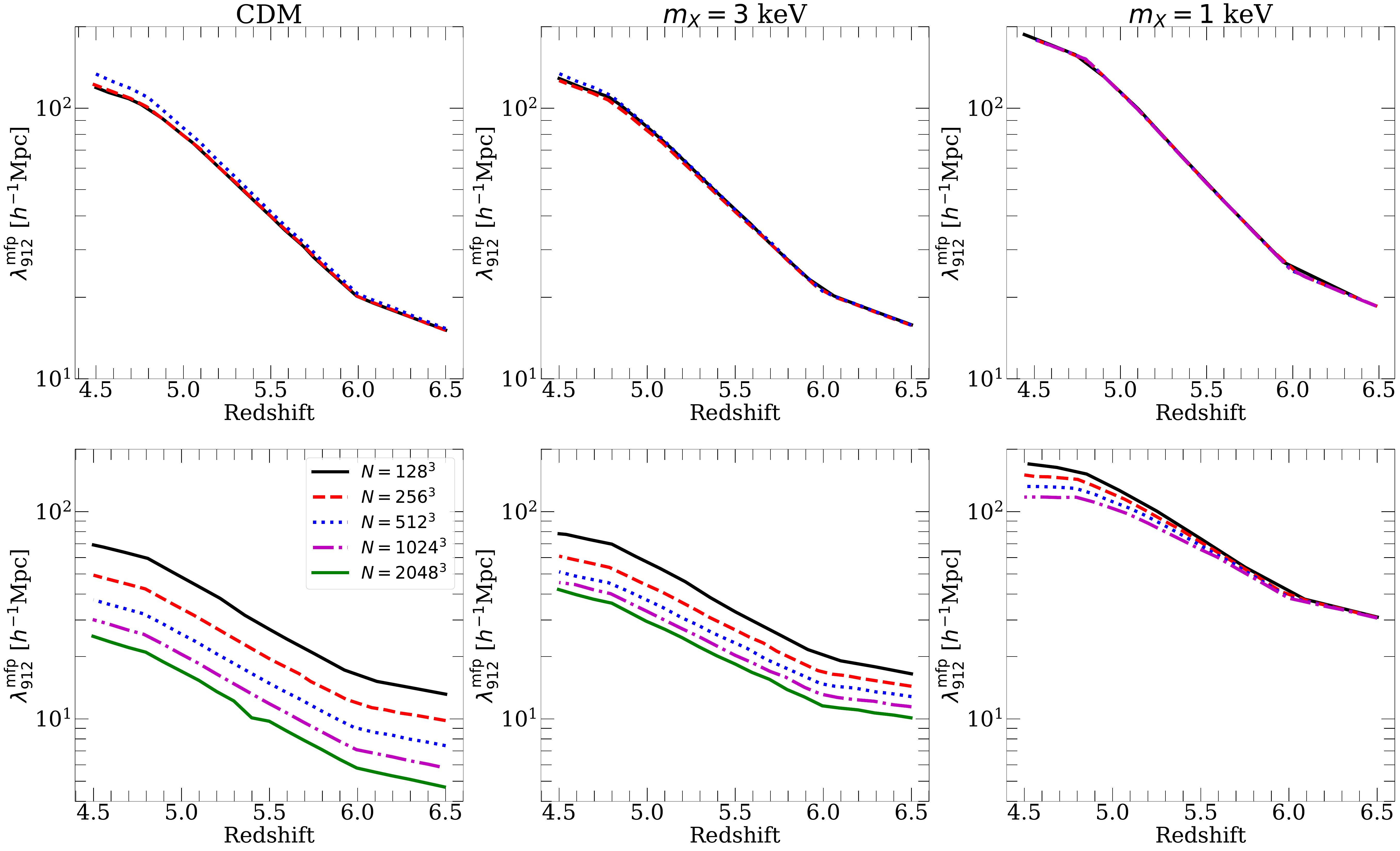}
    \caption{
    Numerical convergence test for our simulation suite.  We show the MFP in ionized gas in the mean-density volume for each of our three DM scenarios (from left to right, CDM, $m_X = 3$ keV and $m_X = 1$ keV), in the relaxed and un-relaxed limits (top and bottom rows respectively).  We tested several resolution levels, the highest in each panel corresponding to the production resolution for that combination of DM model and relaxation state.  The top row shows that our production run resolution of $N = 1024^3$ is more than sufficient for convergence in the relaxed limit.  The bottom row shows that convergence requirements are less stringent in cosmologies with a larger free streaming scale (lower $m_X$).  The lack of convergence in the CDM run highlights the importance of small-scale power in setting the MFP in the un-relaxed limit.  Given that our WDM runs are better converged than our CDM runs, our main results likely underestimate differences in the global MFP between these two cosmologies. We emphasize, however, that our runs do not include any pre-heating by X-ray sources, which would diminish these differences as well.}   
    \label{fig:mfp_chardin_conv}
\end{figure}

We tested the numerical convergence of the MFP estimation (Eq.~\ref{eq:chardin_mfp}) in our simulations in different DM cosmologies in the relaxed and un-relaxed limits.  The test simulations were run in a box with $L=2 h^{-1}$Mpc, with no DC mode.   In the relaxed limit, our fiducial $\Gamma_{\rm HI}$ history was applied. Figure~\ref{fig:mfp_chardin_conv} shows the MFP in the relaxed (top row) and un-relaxed (bottom row) limits for different resolution choices, indicated in the legend.  (Note that we did not run a case with $N=2048^3$ in the un-relaxed $m_X = 1$ keV scenario, so only four curves appear in that panel.)  The left, middle, and right columns show convergence tests for CDM, $m_X = 3$ keV, and $m_X = 1$ keV, respectively.  We see that in the relaxed limit, our choice of $N = 1024^3$ is more than sufficient for numerical convergence for all three DM models.  

In contrast, convergence is extremely difficult in the un-relaxed limit, which has already been noted in \cite{Emberson2013} and \cite{DAloisio2020}. Recall that the un-relaxed production run resolutions for our CDM and $m_X = 3$ keV simulations are both $N=2048^3$, while our $m_X = 1$ keV simulations were run with $N=1024^3$.   The key takeaway from the bottom row of Figure \ref{fig:mfp_chardin_conv} is that the degree of convergence appears to improve as $m_X$ decreases. For example, at $z=6$, the main redshift of interest for this work, the percent differences between adjacent curves for CDM, starting at the lowest resolutions, are 37\%, 32\%, 27\%, and 23\%. For $m_X = 3$ keV, they are 21\%, 14\%, 13\%, and 12\%. For $m_X = 1$ keV, they are 5\%, 2\%, and 1\%.  Hence, even with $N=1024^3$, our production run with $m_X=1$ keV is likely better converged than our CDM run (which was run with $N=2048^3$), justifying our use of a lower resolution for the former.  This feature owes to the intrinsic lack of small-scale power in the WDM cosmology with $m_X =1$ keV.  Another takeaway here is that our main results likely underestimate the LyC opacity in the CDM model, and therefore underestimate the differences in MFP between the WDM and CDM models. Note, however, that our simulations do not include the effect of pre-heating the IGM by the first X-ray sources.  This heating would raise the Jeans filtering mass, smoothing out the smallest structures present in the CDM cosmology.  Ref \cite{DAloisio2020} found that this could lead to as much as a factor of 2 decrease in the IGM clumping factor, which would act in the direction of diminishing differences in the MFP between the CDM and WDM models in the un-relaxed limit.

\section{Testing the relaxation ansatz (Eq. \ref{eq:kappaev})}
\label{app:patchy}

\begin{figure}
    \centering
    \includegraphics[scale=0.19]{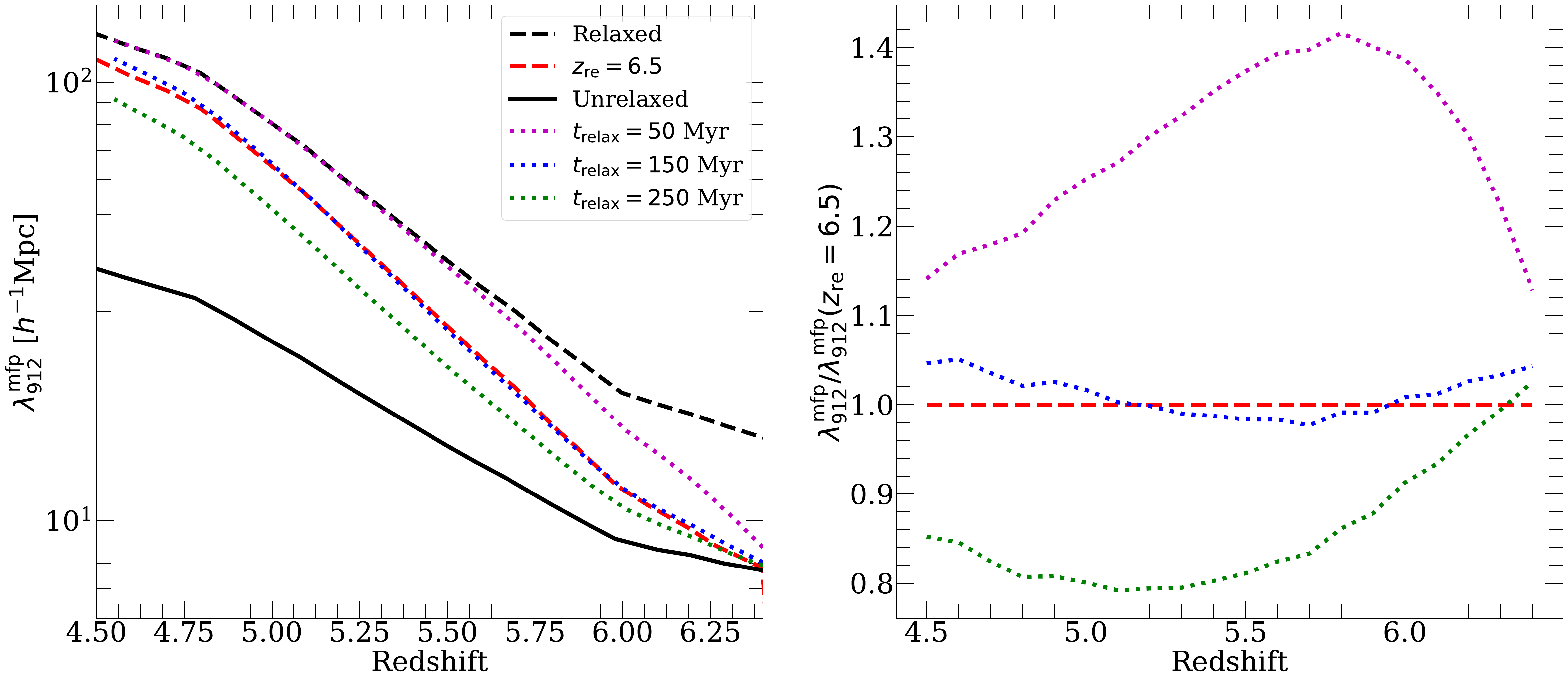}
    \caption{Test of the relaxation ansatz (Eq. \ref{eq:kappaev}).  {\bf Left}: MFP for our simulation with $z_{\rm re} = 6.5$ (red dashed), the corresponding relaxed and un-relaxed limits (black dashed and solid, respectively) and the result of equation~\ref{eq:kappaev} for several values of $t_{\rm relax}$ (dotted curves).  {\bf Right}: ratio between the MFP from the $z_{\rm re} = 6.5$ simulation and the results of equation~\ref{eq:kappaev}.  We see agreement to within at least 10\% between equation~\ref{eq:kappaev} and the simulation for $t_{\rm relax} = 150$ Myr, while the other choices of $t_{\rm relax}$ significantly under or over-shoot the simulation result.  }
    \label{fig:relaxation_test}
\end{figure}

In this section we examine the accuracy of the simple relaxation ansatz given by equation~\ref{eq:kappaev}.  To test this, we ran a hydrodynamic simulation in a CDM cosmology with $N = 512^3$ which was flash re-ionized at $z_{\rm re} = 6.5$ using our fiducial $\Gamma_{\rm HI}$ prescription. 
For comparison, we then ran $N = 512^3$ relaxed and un-relaxed limit runs, and plugged these into equation~\ref{eq:kappaev} for several values of $t_{\rm relax}$.  The left panel in Figure~\ref{fig:relaxation_test} shows the MFP for the simulation with $z_{\rm re} = 6.5$ (red dashed), the relaxed and un-relaxed limits (black dashed and black solid respectively) and the results of evaluating equation~\ref{eq:kappaev} $t_{\rm relax} = 50$, $150$, and $250$ Myr (dotted curves). The right panel shows the fractional difference with the $z_{\rm re} = 6.5$ simulation for each value of $t_{\rm relax}$.  We see that for our fiducial choice of $t_{\rm relax} = 150$ Myr, the ansatz reproduces the simulation result to within at most 10\% at $4.5 < z < 6.5$.  Higher (lower) values of $t_{\rm relax}$ produce MFPs that are distinctly too short (long) compared to the simulation.  This test validates both our simple relaxation ansatz and our our fiducial choice of $t_{\rm relax} = 150$ Myr.   

\acknowledgments

We are grateful to Hy Trac for providing his RadHydro code, which was used to run all hydrodynamic simulations in this paper. We also thank Matt McQuinn and George Becker for helpful discussions, Daniel Gilman for providing his lensing constraints (Fig. 1), and Simeon Bird for his help running MP-Gadget.  A.D.'s group is supported by NASA 19-ATP19-0191 and
NSF AST-2045600.  VI is supported by the Kavli Foundation. All computations were made possible by NSF XSEDE allocation TG-PHY210041 and the NASA HEC Program through the NAS Division at Ames Research
Center.

\bibliography{references}
\bibliographystyle{JHEP}

\end{document}